\documentclass{article}

\usepackage[british]{babel}
\usepackage[useregional]{datetime2}
\DTMlangsetup[en-GB]{showdayofmonth=false}
\usepackage[numbers, sort]{natbib}
\usepackage{array}
\usepackage{booktabs}

\usepackage[a4paper,top=2cm,bottom=2cm,left=2.5cm,right=2.5cm,marginparwidth=1.75cm]{geometry}
\usepackage{amsmath}
\usepackage{amsfonts}
\usepackage{amssymb}
\usepackage{mathtools}
\usepackage{graphicx}
\usepackage{subcaption}
\usepackage{booktabs}
\usepackage{tikz}
\usetikzlibrary{arrows.meta, decorations.pathreplacing}
\usepackage{xcolor}
\definecolor{noarb}{RGB}{46,139,87}
\definecolor{arbprofit}{RGB}{200,40,40}
\definecolor{gascost}{RGB}{38,139,210}
\definecolor{axgray}{RGB}{110,110,110}
\definecolor{bgfill}{RGB}{253,246,227}
\definecolor{strikefill}{RGB}{255,245,200}
\definecolor{arberA}{RGB}{200,40,40}
\definecolor{arberB}{RGB}{38,139,210}
\definecolor{arberC}{RGB}{113,143,0}
\definecolor{marketprice}{RGB}{181,137,0}
\definecolor{poolprice}{RGB}{88,93,176}
\definecolor{weightchange}{RGB}{191,44,120}
\definecolor{bandfill}{RGB}{200,235,210}
\definecolor{dangerfill}{RGB}{255,215,215}
\definecolor{interpA}{RGB}{200,40,40}
\definecolor{interpB}{RGB}{38,139,210}
\usepackage[colorlinks=true, allcolors=blue]{hyperref}
\usepackage[page]{appendix}

\renewcommand{\vec}[1]{\boldsymbol{\mathbf{#1}}}
\renewcommand{\v}{\vec}

\usepackage{authblk}

\title{\vspace*{-1cm}\textbf{Pools as Portfolios: Observed arbitrage efficiency \& LVR analysis of dynamic weight AMMs}}

\author{Matthew Willetts and Christian Harrington}
\affil{QuantAMM.fi}
\begin{document}
\maketitle

\begin{abstract}

Dynamic-weight AMMs (aka Temporal Function Market Makers, TFMMs) implement algorithmic asset allocation, analogous to index or smart beta funds, by continuously updating pools' weights.
A strategy updates target weights over time, and arbitrageurs trade the pool back toward those weights.
This creates a sequence of small, predictable mispricings that grow until taken, effectively executing rebalances as a series of Dutch reverse auctions.
Prior theoretical and simulation work \cite{rvr} predicted that this mechanism could outperform CEX-style rebalancing.
We test that claim on two live pools on the QuantAMM protocol, one on Ethereum mainnet and one on Base, across two short rebalancing windows six months apart (July 2025 and January 2026).
We perform block-level arbitrage analysis, and then measure long term outcomes using Loss-vs-Rebalancing (LVR) and Rebalancing-vs-Rebalancing (RVR) benchmarks.
On mainnet, rebalancing becomes markedly more efficient over time (more frequent arbitrage trades with lower value extracted per trade), reaching performance comparable to or better than CEX-based models.
On Base, rebalancing persists even when per-trade extraction is near (or below) zero, consistent with routing-driven execution, and achieves efficiencies that meet or exceed standard ``perfect rebalancing'' LVR baselines.
These results demonstrate dynamic-weight AMMs as a competitive execution layer for tokenised funds, with superior performance on L2s where routing and lower data costs compress arbitrage spreads.

\end{abstract}

\section{Introduction}

Temporal Function Market Making (TFMM) pools use AMM mechanisms not for core liquidity providing, but instead as a way to run a time-varying portfolio.
They are geometric mean market maker (G3M) pools where the weights change over time.
The weights \emph{are} the portfolio vector, and when a pool's holdings are not in alignment with its weights there is an arbitrage opportunity that, when taken, rebalances the pool to its desired holdings.

At a chosen cadence the pool's quantitative asset management strategy can be run, outputting new target weights (e.g. given recent market data, increase holdings of asset A and reduce holdings of asset B).
Then, analogous to how, when trading on an order book, dividing up a large trade improves execution quality by reducing market impact, the weight change for a TFMM pool is best spread out over time, so the pool interpolated towards the new target holdings.

Arbitrageurs just see a G3M pool with weights that shift.
The pool's quoted prices drift away from the external market, creating a small arbitrage opportunity that grows over time.
Without arbitrageurs needing to know it, this \emph{is} a Dutch reverse auction~\cite{dutchreverse}: the pool implicitly offers a rising payment for rebalancing execution, starting at zero after each trade and growing with allocation drift, until an arbitrageur accepts by striking.
The arbitrageur who acts first captures their profit; after the trade, the pool's reserves snap to market-consistent values, and the cycle repeats until the interpolation is complete.

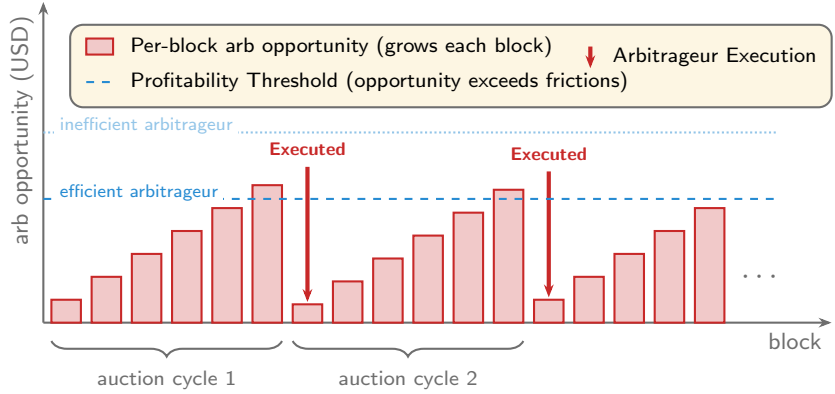
\begin{figure}[t]
\centering
\scalebox{0.98}{\begin{tikzpicture}[
  every node/.style={font=\sffamily},
  >={Stealth[length=5pt]},
  xscale=0.72, yscale=0.62
]
\def\ox{0.8}
\def\oy{0.5}
\draw[->, thick, axgray] (\ox,\oy) -- (15.5,\oy)
  node[below left, font=\sffamily\small] {block};
\draw[->, thick, axgray] (\ox,\oy) -- (\ox,7.5)
  node[rotate=90, anchor=south, font=\sffamily\small, pos=0.6]
  {arb opportunity (USD)};
\def\bw{0.55}
\foreach \i/\h in {1/0.5, 2/1.0, 3/1.5, 4/2.0, 5/2.5, 6/3.0} {
  \pgfmathsetmacro{\xl}{\ox + 0.15 + (\i-1)*0.75}
  \fill[arbprofit!25] (\xl, \oy) rectangle (\xl+\bw, \oy+\h);
  \draw[arbprofit, thick] (\xl, \oy) rectangle (\xl+\bw, \oy+\h);
}
\pgfmathsetmacro{\strikex}{\ox + 0.15 + 6*0.75 + \bw/2}
\draw[arbprofit, ultra thick, ->] (\strikex, 3.9) -- (\strikex, 0.95);
\node[arbprofit, font=\sffamily\scriptsize\bfseries, above] at (\strikex, 3.95) {Executed};
\foreach \i/\h in {1/0.4, 2/0.9, 3/1.4, 4/1.9, 5/2.4, 6/2.9} {
  \pgfmathsetmacro{\xl}{\ox + 0.15 + (\i+6-1)*0.75}
  \fill[arbprofit!25] (\xl, \oy) rectangle (\xl+\bw, \oy+\h);
  \draw[arbprofit, thick] (\xl, \oy) rectangle (\xl+\bw, \oy+\h);
}
\pgfmathsetmacro{\strikexB}{\ox + 0.15 + 12*0.75 + \bw/2}
\draw[arbprofit, ultra thick, ->] (\strikexB, 3.8) -- (\strikexB, 1.05);
\node[arbprofit, font=\sffamily\scriptsize\bfseries, above] at (\strikexB, 3.85) {Executed};
\foreach \i/\h in {1/0.5, 2/1.0, 3/1.5, 4/2.0, 5/2.5} {
  \pgfmathsetmacro{\xl}{\ox + 0.15 + (\i+12-1)*0.75}
  \fill[arbprofit!25] (\xl, \oy) rectangle (\xl+\bw, \oy+\h);
  \draw[arbprofit, thick] (\xl, \oy) rectangle (\xl+\bw, \oy+\h);
}
\node[axgray, font=\sffamily\large] at (14.2, 1.5) {$\cdots$};
\pgfmathsetmacro{\gaslow}{0.5 + 2.7}
\pgfmathsetmacro{\gashigh}{0.5 + 4.15}
\draw[gascost, thick, dashed] (\ox, \gaslow) -- (14.5, \gaslow);
\node[gascost, font=\sffamily\scriptsize, fill=white, inner sep=2pt, anchor=west]
  at (\ox+0.2, \gaslow+0.15) {efficient arbitrageur};
\draw[gascost!50, thick, densely dotted] (\ox, \gashigh) -- (14.5, \gashigh);
\node[gascost!50, font=\sffamily\scriptsize, fill=white, inner sep=2pt, anchor=west]
  at (\ox+0.2, \gashigh+0.15) {inefficient arbitrageur};
\pgfmathsetmacro{\cycAl}{\ox + 0.15}
\pgfmathsetmacro{\cycAr}{\ox + 0.15 + 5*0.75 + \bw}
\pgfmathsetmacro{\cycBl}{\ox + 0.15 + 6*0.75}
\pgfmathsetmacro{\cycBr}{\ox + 0.15 + 11*0.75 + \bw}
\draw[decorate, decoration={brace, amplitude=6pt, mirror}, thick, axgray]
  (\cycAl, 0.1) -- (\cycAr, 0.1);
\node[axgray, font=\sffamily\footnotesize, below=8pt] at ({(\cycAl+\cycAr)/2}, 0.1)
  {auction cycle 1};
\draw[decorate, decoration={brace, amplitude=6pt, mirror}, thick, axgray]
  (\cycBl, 0.1) -- (\cycBr, 0.1);
\node[axgray, font=\sffamily\footnotesize, below=8pt] at ({(\cycBl+\cycBr)/2}, 0.1)
  {auction cycle 2};
\begin{scope}[shift={(1.5, 5.5)}]
  \draw[rounded corners=5pt, thick, axgray, fill=bgfill]
    (-0.2,-0.3) rectangle (14.0, 1.5);
  \fill[arbprofit!25] (0.1, 0.85) rectangle (0.6, 1.2);
  \draw[arbprofit, thick] (0.1, 0.85) rectangle (0.6, 1.2);
  \node[font=\sffamily\footnotesize, anchor=west] at (0.75, 1.025)
    {Per-block arb opportunity (grows each block)};
  \draw[gascost, thick, dashed] (0.1, 0.25) -- (0.6, 0.25);
  \node[font=\sffamily\footnotesize, anchor=west] at (0.75, 0.25)
    {Profitability Threshold (opportunity exceeds frictions)};
  \draw[arbprofit, ultra thick, ->] (9.5, 1.025) -- (9.5, 0.55);
  \node[font=\sffamily\footnotesize, anchor=west] at (9.75, 0.8)
    {Arbitrageur Execution};
\end{scope}
\end{tikzpicture}}
\caption{The Dutch reverse auction mechanism. Each block, the arb opportunity (red bars) grows as weight changes accumulate. The dashed lines mark arbitrageur cost thresholds; the trade occurs when the opportunity exceeds the leanest arbitrageur's cost. After the trade the cycle resets.
}
\label{fig:auction_schematic}
\vspace{-1em}
\end{figure}

The competitive arbitrageur ecosystem is the pool's execution engine; the question is how much this execution costs.
Figure~\ref{fig:auction_schematic} illustrates this mechanism schematically.
Previous work has characterised this mechanism theoretically~\cite{willetts2024closedform, willetts2024optimalrebalancingdynamicamms} and in simulation.
The Loss-versus-Rebalancing (LVR) framework~\cite{lvr, lvr_arb} and its extension Rebalancing-versus-Rebalancing (RVR)~\cite{rvr} provide benchmarks for the cost of AMM-based rebalancing, and the MEV literature~\cite{daian2020flashboys} establishes that on-chain arbitrage is mediated by a competitive ecosystem of searchers.

First we give a per-block empirical analysis of arb dynamics on live dynamic-weight AMM pools on the QuantAMM protocol, the first such study, revealing auction compression in the a six-month period from July 2025 to early 2026.
We identify \emph{incidental routing}---rebalancing that occurs as a side effect of ecosystem-wide DEX routing at near-zero cost to the pool.
We then benchmark the pools' properties against LVR and RVR.

\section{Background}

\paragraph{G3M pools}

A geometric mean market maker (G3M)~\cite{evansG3Ms, balancer} with $N$ tokens has trading function $\prod_{i=1}^N R_i^{w_i}= k$, where $\v R =\{R_i\}$ is the $N$-vector of currently-held reserves of each token and the weights $\v w =\{w_i\}$, $\sum_{i=1}^N w_i = 1$ and $\forall i\,\, 0\leq w_i<1$, control how much value is stored in each token.
The pool's \emph{actual} value allocation for token $i$ is $\theta_i = \frac{R_i\, p_i}{\sum_j R_j\, p_j}$.
A fundamental property of G3M pools is that $\theta_i = w_i$ only when the pool quotes market prices~\cite{angeris2020does}.
A Temporal Function Market Maker (TFMM)~\cite{tfmm, tfmm_litepaper} is a G3M where the weights change with time according to a set process.
The trading function becomes $\prod_{i=1}^N R_i^{w_i(t)}= k(t)$.
Within a block weights are constant (i.e. $t$ is the discrete blocknumber).

When the pool drifts from market prices (due to weight changes or external price movements), $\theta_i \neq w_i$, until the pool gets arbed.
A pool will not necessarily get arbed immediately, the pool's own fees mean the price discrepancy has be be `large enough`, and other real world frictions (gas cost, the efficiency of other venues the arbitrageur might use to get into their chosen numeraire at the end of the arbitrage trade, time delays inherent to blockchain systems) increase the effective `no-arb region` further.

\begin{figure}[p]
\centering
\begin{subfigure}[t]{\textwidth}
\centering
\resizebox{0.85\textwidth}{!}{\begin{tikzpicture}[
  every node/.style={font=\sffamily},
  >={Stealth[length=5pt]}
]

\begin{scope}[yshift=3cm]
  \node[font=\sffamily\bfseries, anchor=west] at (0, 1.5)
    {(a) No arbitrage --- $m_p$ inside band};

  \draw[->, thick, axgray] (0.5,0) -- (14.5,0);

  \fill[bandfill, rounded corners=3pt] (3,-0.4) rectangle (11,0.4);
  \draw[noarb, very thick, rounded corners=3pt] (3,-0.4) rectangle (11,0.4);

  \draw[noarb, thick] (3, -0.4) -- (3, -0.8);
  \draw[noarb, thick] (11,-0.4) -- (11,-0.8);
  \node[noarb, font=\sffamily\small, below] at (3,-0.8) {$\gamma\, m_u$};
  \node[noarb, font=\sffamily\small, below] at (11,-0.8) {$\gamma^{-1} m_u$};

  \fill[poolprice] (7,0) circle (5pt);
  \node[poolprice, font=\sffamily\small, above=5pt] at (7,0.4) {$m_u$};

  \fill[marketprice] (8.8,0) circle (5pt);
  \node[marketprice, font=\sffamily\small, above=5pt] at (8.8,0.4) {$m_p$};

  \node[noarb, font=\sffamily\small\itshape] at (7,-1.7)
    {Market price within band $\Rightarrow$ no profitable trade exists};

  \node[noarb, font=\sffamily\Large\bfseries] at (13.5, 0) {\checkmark};
\end{scope}

\begin{scope}[yshift=-1cm]
  \node[font=\sffamily\bfseries, anchor=west] at (0, 1.5)
    {(b) Arbitrage opportunity --- $m_p$ outside band};

  \draw[->, thick, axgray] (0.5,0) -- (14.5,0);

  \fill[bandfill, rounded corners=3pt] (3,-0.4) rectangle (11,0.4);
  \draw[noarb, very thick, rounded corners=3pt] (3,-0.4) rectangle (11,0.4);

  \draw[noarb, thick] (3, -0.4) -- (3, -0.8);
  \draw[noarb, thick] (11,-0.4) -- (11,-0.8);
  \node[noarb, font=\sffamily\small, below] at (3,-0.8) {$\gamma\, m_u$};
  \node[noarb, font=\sffamily\small, below] at (11,-0.8) {$\gamma^{-1} m_u$};

  \fill[poolprice] (7,0) circle (5pt);
  \node[poolprice, font=\sffamily\small, above=5pt] at (7,0.4) {$m_u$};

  \fill[marketprice] (12.8,0) circle (5pt);
  \node[marketprice, font=\sffamily\small, above=5pt] at (12.8,0.4) {$m_p$};

  \fill[dangerfill, opacity=0.6] (11,-0.4) rectangle (12.8,0.4);

  \draw[arbprofit, ultra thick, <->] (11, -1.4) -- (12.8, -1.4);
  \node[arbprofit, font=\sffamily\small, below=2pt] at (11.9, -1.4)
    {profit $\propto$ gap};

  \node[arbprofit, font=\sffamily\small\itshape] at (7,-2.2)
    {Market price outside band $\Rightarrow$ profitable arbitrage trade exists};

  \node[arbprofit, font=\sffamily\Large\bfseries] at (14.0, 0) {!};
\end{scope}

\end{tikzpicture}}
\caption{The no-arbitrage region for a two token pool.
(a) Market price ~$m_p$ inside the band $[\gamma m_u, \gamma^{-1}m_u]$ (where $m_u$ is the price quoted by the pool for a marginal trade): no profitable trade.
(b) Market price ~$m_p$ outside the band: profit proportional to gap. An optimal arbitrageur places $m_p$ at the boundary of the \emph{new} band, leaving a residual gap $\approx$ fee (see also Figure~\ref{fig:boundary_targeting}).}
\label{fig:noarb_schematic}
\end{subfigure}

\vspace{2mm}

\begin{subfigure}[t]{\textwidth}
\centering
\resizebox{0.85\textwidth}{!}{\begin{tikzpicture}[
  every node/.style={font=\sffamily},
  >={Stealth[length=5pt]}
]

\newcommand{\drawrow}[8]{%
  \begin{scope}[yshift=#1 cm]
    \draw[axgray, thin] (1, 0) -- (11, 0);
    \fill[bandfill, rounded corners=2pt] (#2,-0.28) rectangle (#3,0.28);
    \draw[noarb, thick, rounded corners=2pt] (#2,-0.28) rectangle (#3,0.28);
    \fill[poolprice] (#4, 0) circle (3.5pt);
    \fill[marketprice] (#5, 0) circle (3.5pt);
    \node[font=\sffamily\bfseries\small, anchor=east] at (0.8, 0) {#6};
    \node[font=\sffamily\footnotesize, text=axgray, anchor=west] at (11.3, 0) {#8};
  \end{scope}
}

\drawrow{10}{3.5}{7.5}{5.5}{5.5}{$t=0$}{}{equilibrium}
\node[poolprice, font=\sffamily\tiny, above=4pt] at (5.5, 10.28) {$m_u$};
\node[marketprice, font=\sffamily\tiny, below=4pt] at (5.5, 9.72) {$m_p$};

\drawrow{8.5}{3.9}{7.9}{5.9}{5.5}{$t=1$}{}{band shifts}
\draw[weightchange, thick, ->, shorten >=1pt] (5.5, 9.0) -- (5.9, 9.0)
  node[above, midway, font=\sffamily\tiny] {$\Delta w$};

\drawrow{7}{4.3}{8.3}{6.3}{5.5}{$t=2$}{}{gap growing}
\draw[weightchange, thick, ->, shorten >=1pt] (5.9, 7.4) -- (6.3, 7.4)
  node[above, midway, font=\sffamily\tiny] {$\Delta w$};

\drawrow{5.5}{5.8}{9.8}{7.8}{5.5}{$t=k$}{}{arb exists!}
\fill[dangerfill, opacity=0.5, rounded corners=1pt] (5.5, 5.22) rectangle (5.8, 5.78);
\fill[marketprice] (5.5, 5.5) circle (3.5pt);
\draw[arbprofit, very thick, <->] (5.5, 5.05) -- (5.8, 5.05);
\node[arbprofit, font=\sffamily\tiny, below=1pt] at (5.65, 5.05) {gap};

\drawrow{4}{6.8}{10.8}{8.8}{5.5}{$t\!=\!k\!+\!1$}{}{gap widens}
\fill[dangerfill, opacity=0.5, rounded corners=1pt] (5.5, 3.72) rectangle (6.8, 4.28);
\fill[marketprice] (5.5, 4) circle (3.5pt);
\draw[arbprofit, very thick, <->] (5.5, 3.55) -- (6.8, 3.55);
\node[arbprofit, font=\sffamily\tiny, below=1pt] at (6.15, 3.55) {larger gap};

\drawrow{2}{5.5}{9.5}{7.5}{5.5}{$t\!=\!k\!+\!2$}{}{post-arb reset}
\draw[arbprofit, ultra thick, ->, rounded corners=3pt]
  (8.8, 3.2) -- (8.8, 2.7) -- (7.5, 2.7) -- (7.5, 2.35);
\node[arbprofit, font=\sffamily\footnotesize\bfseries] at (9.7, 3.0) {arb trade!};
\node[marketprice, font=\sffamily\tiny, below=3pt] at (5.5, 1.72) {$m_p$};
\node[poolprice, font=\sffamily\tiny, below=3pt] at (6.5, 1.72) {$m_u$};
\draw[axgray, thick, <->] (5.5, 1.2) -- (6.5, 1.2);
\node[axgray, font=\sffamily\tiny, below=1pt] at (6.0, 1.2) {gap $\approx$ fee};

\draw[weightchange, ultra thick, ->, dashed, rounded corners=6pt]
  (13.5, 2.0) -- (14.5, 2.0) -- (14.5, 10.0) -- (13.5, 10.0);
\node[weightchange, font=\sffamily\footnotesize, rotate=90, anchor=south]
  at (15.0, 6) {cycle repeats};

\end{tikzpicture}}
\caption{Weight changes shift the no-arb band block-by-block.
$m_u$ (violet) and the band (green) drift rightward while $m_p$ (gold) stays fixed.
By $t\!=\!k$, $m_p$ exits the band; the gap widens until an arbitrageur trades at $t\!=\!k\!+\!2$, resetting $m_u$ so that $m_p$ sits at the new boundary. The dashed arrow marks the repeating cycle.}
\label{fig:band_drift}
\end{subfigure}

\caption{No-arb band mechanics and the weight-driven auction cycle. 
}
\label{fig:noarb_and_drift}
\end{figure}

\paragraph{From weight changes to arbitrage}
\label{sec:weight_to_arb}

As the update rule shifts target weights each block, the pool's quoted marginal price $m_u$ moves and the no-arb band moves with it.
If external prices are relatively stable, the band ``walks away'' from the market price (Figure~\ref{fig:band_drift}).

At some block $t = k$, the market price exits the band and an arb opportunity appears.
The gap grows each block as weights continue to interpolate, until an arbitrageur trades.
The arbitrageur trades just enough to shift $m_u$ so that $m_p$ lands by the boundary of the new band, leaving a residual gap of approximately one fee.
The cycle then repeats, a series of instant runoff Dutch reverse auction, where each block of weight interpolation adds a small increment to the arb opportunity, and the first arbitrageur whose cost falls below the accumulated surplus captures it.

\paragraph{Weight interpolation reduces rebalancing cost}
\label{para:interpolation}

Rebalancing by changing weights over time preserves a G3M pool's value to first order: $V' = V + \mathcal{O}(\delta w^2)$.
This quadratic leading order dependence means total rebalancing cost reduces with the size of constituent weight changes~\cite{tfmm_litepaper}: splitting a single large weight change $\Delta w$ into $N$ small steps of $\Delta w / N$ reduces the total arb cost from $(\Delta w)^2$ to $(\Delta w)^2 / N$.
In the limit of continuous interpolation ($N \to \infty$), the theoretical cost vanishes entirely\cite{tfmm_litepaper}.

QuantAMM pools make use of this by interpolating weights gradually over many blocks rather than making discrete jumps~\cite{willetts2024optimalrebalancingdynamicamms}.
The rebalancing costs observed in this paper reflect pools already operating in the many-small-steps regime; costs would be much higher if weights changed in one go.

\newpage
\section{Experimental setup}

\paragraph{Pools and data}

We study two QuantAMM pools on different chains, at different scales, each observed at two points six months apart:

\begin{table}[h]
\centering
\begin{tabular}{lll}
\toprule
& \textbf{Safe Haven} & \textbf{Base Macro} \\
\midrule
Chain & Ethereum mainnet & Base L2 \\
TVL & ${\sim}$\$300k & ${\sim}$\$50k \\
Launch Date & May 2025 & May 2025 \\
Assets & BTC / PAXG / USDC & AERO / BTC / ETH / USDC \\
Swap fee & 0.3\% & 0.3\% \\
Protocol fee & 50\% of swap fee & 50\% of swap fee \\
Block time & 12s & 2s \\
MEV infrastructure & PBS (Flashbots) & Sequencer (FIFO) \\
\bottomrule
\end{tabular}
\caption{The two pools share fee structure and protocol design but differ in TVL (${\sim}50\times$), chain architecture, and block frequency ($6\times$).}
\label{tab:pools}
\end{table}

For each pool and window, we scrape per-block reserves, weights, prices, and gas statistics, identifying blocks where pool balances change due to swaps then matching each change in reserves to its on-chain transaction.\footnote{ QuantAMM protocol runs on top of Balancer V3's vault which keeps 50\% of the pool's swap fee, so pools' balance deltas must be scaled appropriately before further calculation.}

\paragraph{Profitability threshold}
\label{sec:threshold}

We construct a time-varying, per-block profitability threshold calibrated against observed arbitrageur costs.
The total cost for an arbitrageur is made up of three components: the EIP-1559 base fee (burned), a priority fee (paid to the block builder), and optionally a direct builder tip~\cite{daian2020flashboys, flashbots}.
We identify \emph{open-market arbitrageurs}---those whose full cost is visible on-chain---and compute the threshold for a block $b$ as
\[
\text{threshold}(b) = \text{baseFee}(b) \times G \times p_\text{ETH} \times \mu \;/\; 10^9,
\]
where $G = 450{,}000$ gas units (median amount for observed arb trades), $p_\text{ETH}$ is the ETH price, and $\mu$ is the calibrated markup (capturing the intensity of competition in the arbitrage market).
 Looking at open-market trades in July in different arbitrageurs, the mean markup is $\mu = 1.50$.

\paragraph{Mispricing}
How to measure how far a pool's mispricing is, in aggregate, to best observe how mispricing builds up over time and is then reduced by an arbitrage trade?
We study multi-token ($N=3$ or $4$) pools here.
We define the \emph{allocation drift} as $\sum_i |\theta_i - w_i|$, which is zero at equilibrium and grows as the pool deviates.
As we will show, this metric produces a clean sawtooth pattern with gradual buildup as the pool drifts, sharp resets when an arbitrageur trades the pool back toward its target allocation, without having to handle or plot $\sim N^2/2$ pairwise price comparisons.

\section{Results}

First we give extremely fine grained analysis of the arbitrage trades done against these pools over two time short periods where pool weights are changing, each a couple of hours, one in July 2025 a few months after launch, and more recently in Jan 2026.
Second, we give broad, aggregate analysis of these pools since their launch, benchmarking against Loss-versus-Rebalancing (LVR~\cite{lvr}) and its more realistic extension Rebalancing versus Rebalancing (RVR~\cite{rvr}) that models the spread and commission fees present when trading on a CEX.

\subsection{Block-level arbitrage analysis}
\label{ssec:block_level}

\paragraph{Auction dynamics on Ethereum mainnet}

\begin{figure}[h]
\centering
\includegraphics[width=\textwidth]{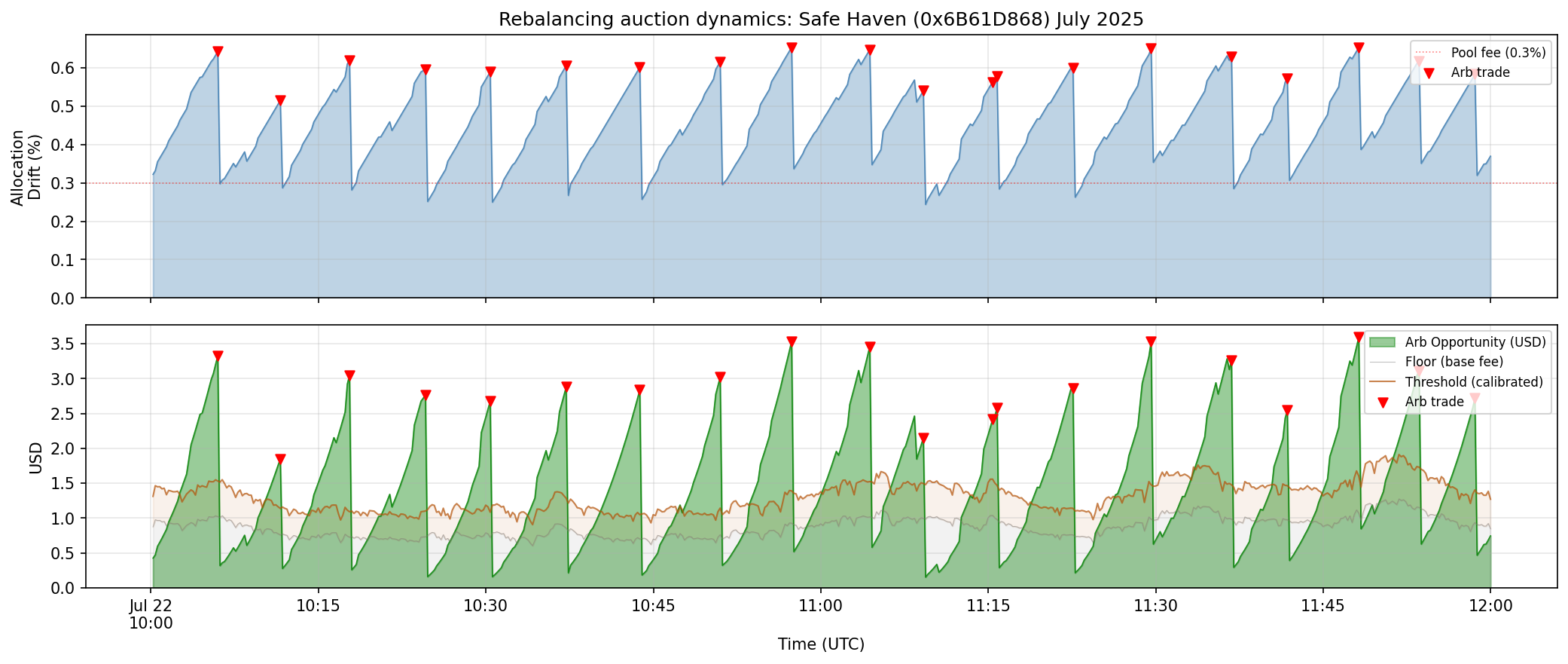}
\caption{Safe Haven pool, July 22 2025 (10:00--12:00 UTC), 606 blocks, 20 arb trades.
\textbf{Top:} Allocation drift sawtooth; dashed line = 0.3\% pool fee.
\textbf{Second:} Per-block arb profit (green) with floor and calibrated thresholds; red triangles = arbitrage trades. All 20 exceed the realistic threshold.}
\label{fig:safe_haven_july}
\end{figure}

In a two-hour window from the Safe Haven pool on July 22 2025 (10:00--12:00 UTC), we observe 20 trades over 606 blocks.
The allocation drift sawtooth is clean (Figure~\ref{fig:safe_haven_july}): drift builds to 0.5--0.65\% between trades, then snaps back as an arbitrageur trades the pool toward its target.
Arbitrage trades occur roughly every 30 blocks (${\sim}$6 minutes).

The closed-form optimal arbitrage trade~\cite{willetts2024closedform} predicts a theoretical maximum extraction of \$58.19 over the window.
arbitrageurs empirically extract \$51.55, an efficiency ratio of 88.6\%.
The gap is explained primarily by conservative trade sizing (Appendix~\ref{app:undertrading}): arbitrageurs consistently execute $\sim60$\% of the optimal trade size while capturing almost all of the available profit, which is natural given that the G3M arb profit function is concave and flat near its maximum (Figure~\ref{fig:boundary_targeting}).

\begin{figure}[h]
\centering
\includegraphics[width=\textwidth]{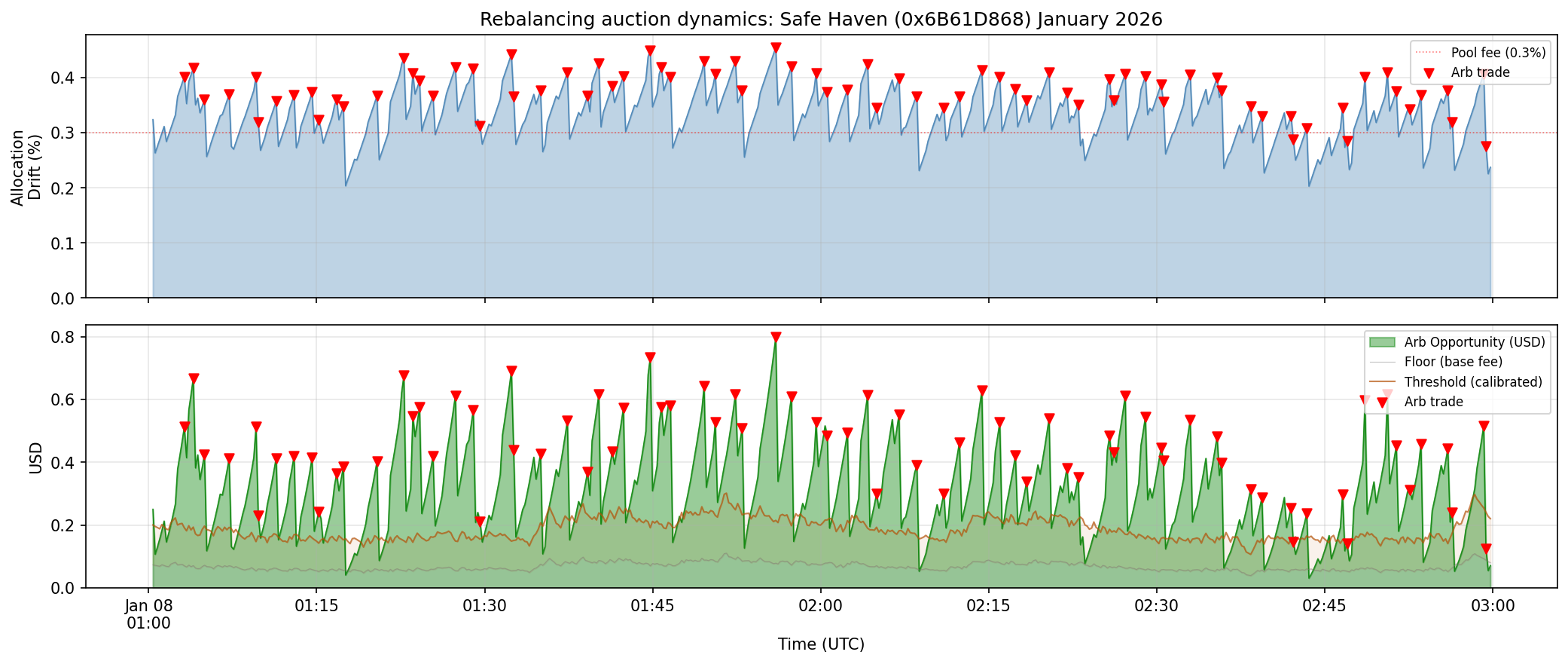}
\caption{Safe Haven pool, January 8 2026 (01:00--03:00 UTC), 591 blocks, 78 trades (${\sim}4\times$ July).
The sawtooth is compressed: drift rarely exceeds 0.45\% and trades occur every ${\sim}$90 seconds.
The y-axis scale is ${\sim}4\times$ smaller than Figure~\ref{fig:safe_haven_july}; 32\% of trades fall below the standalone threshold.
}
\label{fig:safe_haven_jan}
\end{figure}

Six months later, Jan 2026, the same pool looks very different (Figure~\ref{fig:safe_haven_jan}).
The sawtooth is compressed: allocation drift barely exceeds 0.4\% before an arbitrageur trades.
Instead of 20 trades in two hours, we observe 78---roughly one every 90 seconds.
Per-trade empirical extraction falls from \$2.58 to \$0.28, and total empirical extraction drops from \$51.55 to \$22.00 despite nearly $4\times$ more trades (Table~\ref{tab:safe_haven_comparison}).

\begin{table}[h]
\footnotesize %
\centering
\begin{tabular}{lrr}
\toprule
& \textbf{July 2025} & \textbf{January 2026} \\
\midrule
Trades (2-hour window) & 20 & 78 \\
Empirical extraction per trade & \$2.58 & \$0.28 \\
Total empirical extraction & \$51.55 & \$22.00 \\
Theoretical optimum & \$58.19 & \$35.82 \\
Max allocation drift & ${\sim}$0.65\% & ${\sim}$0.45\% \\
Blocks between trades & ${\sim}$30 (${\sim}$6 min) & ${\sim}$7.5 (${\sim}$90 sec) \\
\bottomrule
\end{tabular}
\caption{Safe Haven arb dynamics across two windows. All profit figures are empirical.
}
\label{tab:safe_haven_comparison}
\end{table}

This is the dynamic predicted by the Dutch reverse auction framework~\cite{willetts2024optimalrebalancingdynamicamms}: competition among searchers progressively lowers the arbitrage threshold, the auction clears at a lower price, and the pool benefits.

No retail or organic swap flow was detected in either window: every transaction originates from MEV market participants, identified by contract-mediated execution and/or multi-hop routing through 2+ DEXs.
By January, the routing landscape has shifted: Uniswap V4 accounts for 37\% of all swap events (up from zero), complex MEV bundles with 4+ legs account for 24\% of trades, and the mean all-in cost per transaction fell from \$1.08 to \$0.20, drive by the Fusaka Upgrade in Dec 2025 (see Appendix~\ref{app:tx_structure} for full breakdown).

\begin{figure}[h]
\centering
\includegraphics[width=\textwidth]{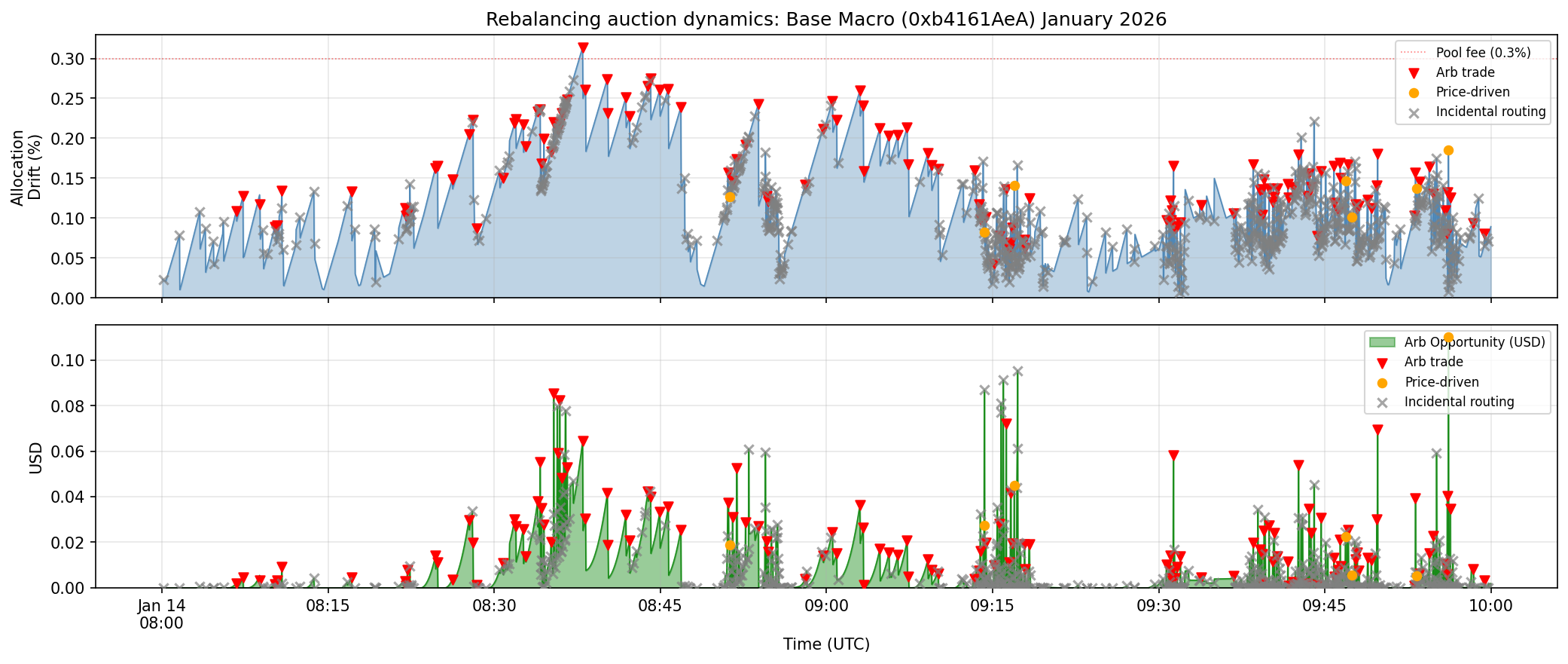}
\caption{Base Macro pool, January 14 2026 (08:00--10:00 UTC). The sawtooth is present but individual resets are tiny.
A small number of arb trades here are driven by changes in external asset prices during this weight interpolation, marked in yellow, where we have found them by taking a `what for' approach: if we can still find an arbitrage opportunity in that moment from the current pool reserves, current market prices, and \emph{stale} weights, then we say the trade is `price drive' and is not a key part of our weight-change-rebalance analysis.
Trades resulting in negative profit for the arbitrageur (a zero time markout) we label as `incidental routing'; these are legs of arb transactions whose profit is generated elsewhere.
}
\label{fig:base_macro_jan}
\end{figure}

\paragraph{Incidental rebalancing on Base L2}

The Base Macro pool provides a contrasting case: it sits on a chain with no PBS infrastructure.
In the January window we observe 202 block-to-block balance changes from 225 on-chain transactions.
Per-trade empirical profit is essentially zero: the mean is \$0.0004, the median is below \$0.001, and 98\% of trades extract less than a cent from the pool, many in fact `extracting' less than zero.
These trades, as a result, often add value to the pool, albeit a tiny amount.

These arbitrageurs are not targeting this pool for value extraction.
The pool serves as a routing venue in complex multi-hop arbitrages across the Base DEX ecosystem.
January transactions average 7 swaps across UniV2, V3, and V4, all to handle what amounts to a \$20--80 trade through this pool.
The arbitrage profit comes from price discrepancies between \emph{other} pools; the QuantAMM pool happens to be one hop in the route, traded at approximately fair market value.
We call this \emph{incidental routing}, examples of which are marked with a grey $\mathsf{x}$ in Figure~\ref{fig:base_macro_jan}.

Yet inventory does shift.
The swaps are directionally correct---arbitrageurs buy whichever token is slightly cheap in the pool and sell whichever is slightly expensive---and the allocation drift still shows its characteristic sawtooth (Figure~\ref{fig:base_macro_jan}), though individual resets are tiny.
The mechanism is straightforward: target weights shift slowly, allocation drift builds, but the resulting mispricing is too small to justify standalone arb after the 0.3\% fee plus gas.
Multi-hop arbitrageurs route through the pool anyway as part of a profitable ecosystem-wide route, and the swap moves reserves back toward target.

\subsection{Benchmarking rebalancing efficiency}
\label{ssec:benchmarks}

\begin{figure}[t]
  \centering

  \hspace{2em}%
  \makebox[0.46\textwidth]{\textbf{LVR}}%
  \hfill
  \makebox[0.46\textwidth]{\textbf{\quad RVR}}%

  \vspace{4pt}

  \begin{minipage}[c]{\textwidth}
  \hspace{0.5em}%
  \makebox[0pt]{\rotatebox[origin=c]{90}{\textbf{\qquad Safe Haven}}\hspace*{2em}}%
  \begin{minipage}[c]{\textwidth}
  \centering
  \begin{subfigure}[t]{0.46\textwidth}
  \includegraphics[width=\textwidth]{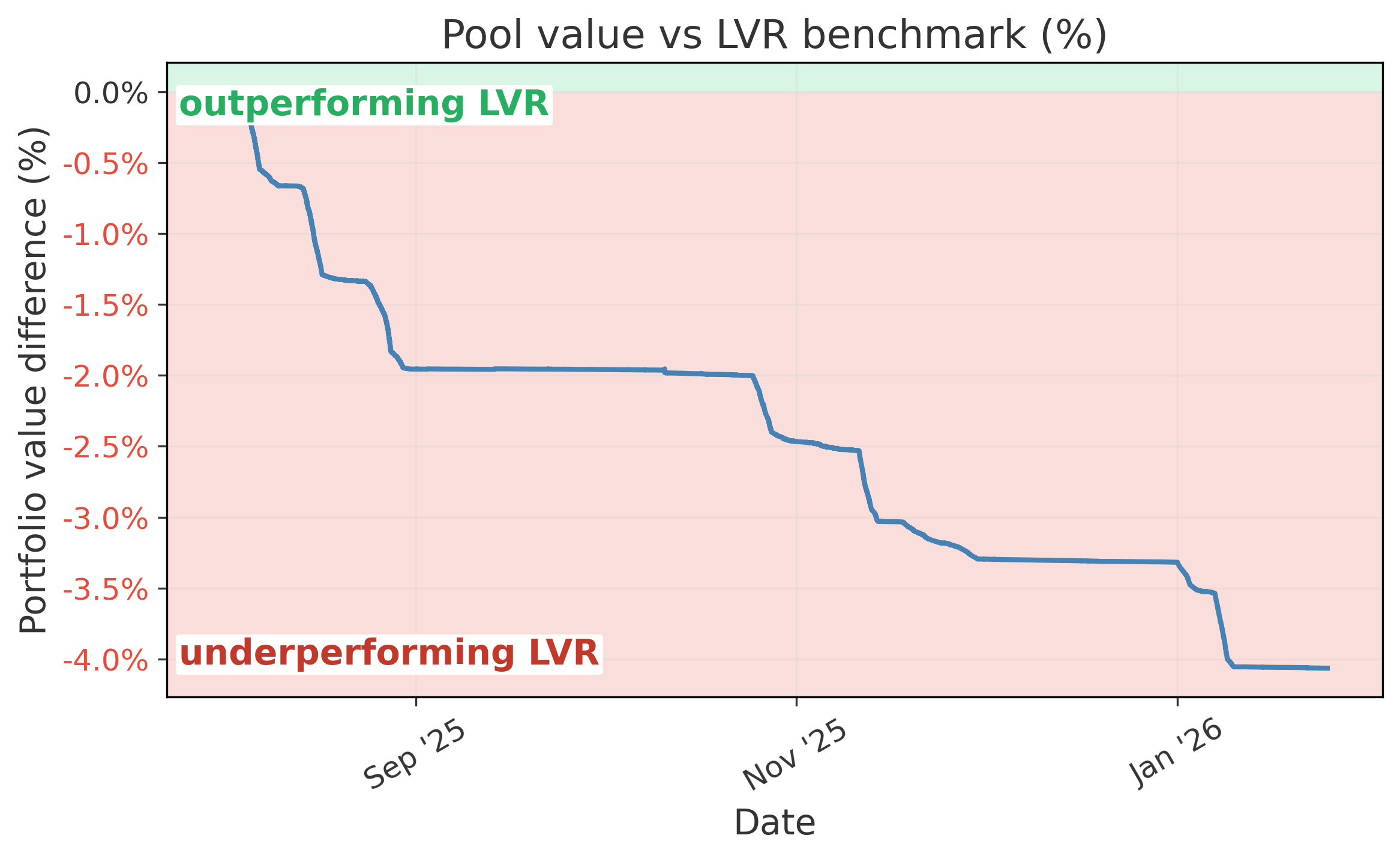}
  \caption{Mainnet, Aug 2025--Jan 2026.}
  \end{subfigure}%
  \hfill
  \begin{subfigure}[t]{0.46\textwidth}
  \includegraphics[width=\textwidth]{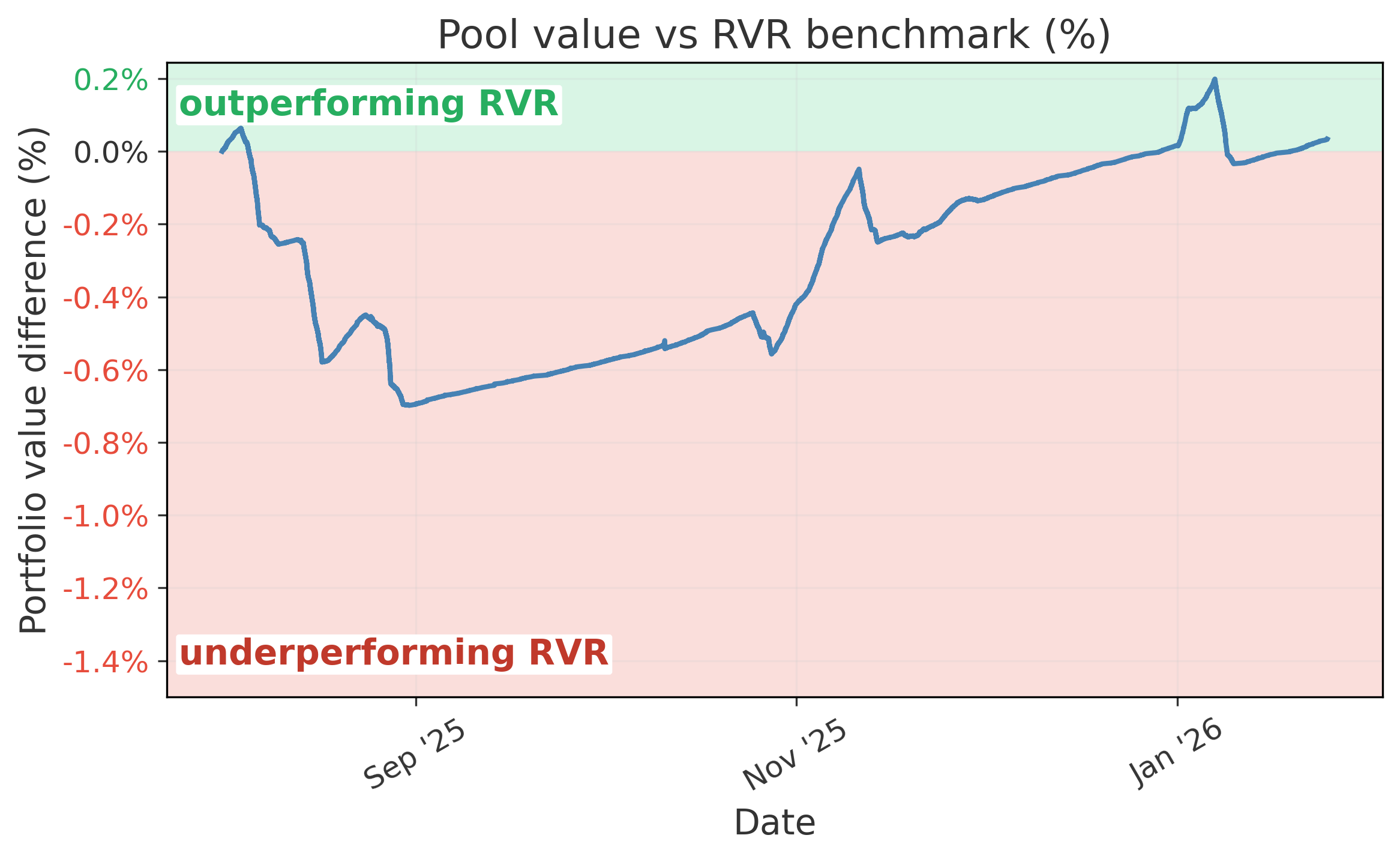}
  \caption{Mainnet, same period.}
  \end{subfigure}
  \end{minipage}
  \end{minipage}

  \vspace{6pt}

  \begin{minipage}[c]{\textwidth}
  \hspace{0.5em}%
  \makebox[0pt]{\rotatebox[origin=c]{90}{\textbf{\qquad Base Macro}}\hspace*{2em}}%
  \begin{minipage}[c]{\textwidth}
  \centering
  \begin{subfigure}[t]{0.46\textwidth}
  \includegraphics[width=\textwidth]{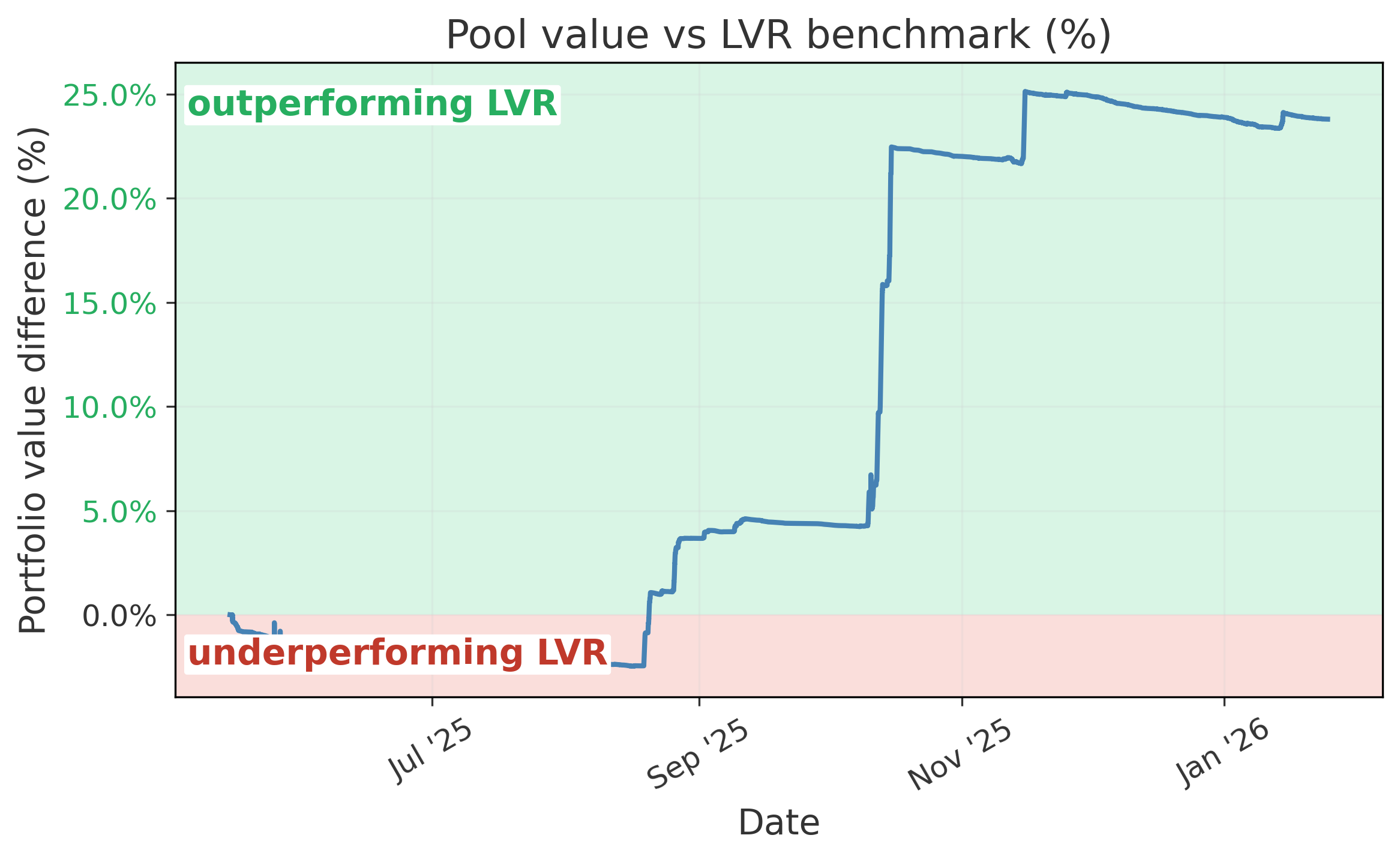}
  \caption{Base L2, May 2025--Jan 2026.}
  \end{subfigure}%
  \hfill
  \begin{subfigure}[t]{0.46\textwidth}
  \includegraphics[width=\textwidth]{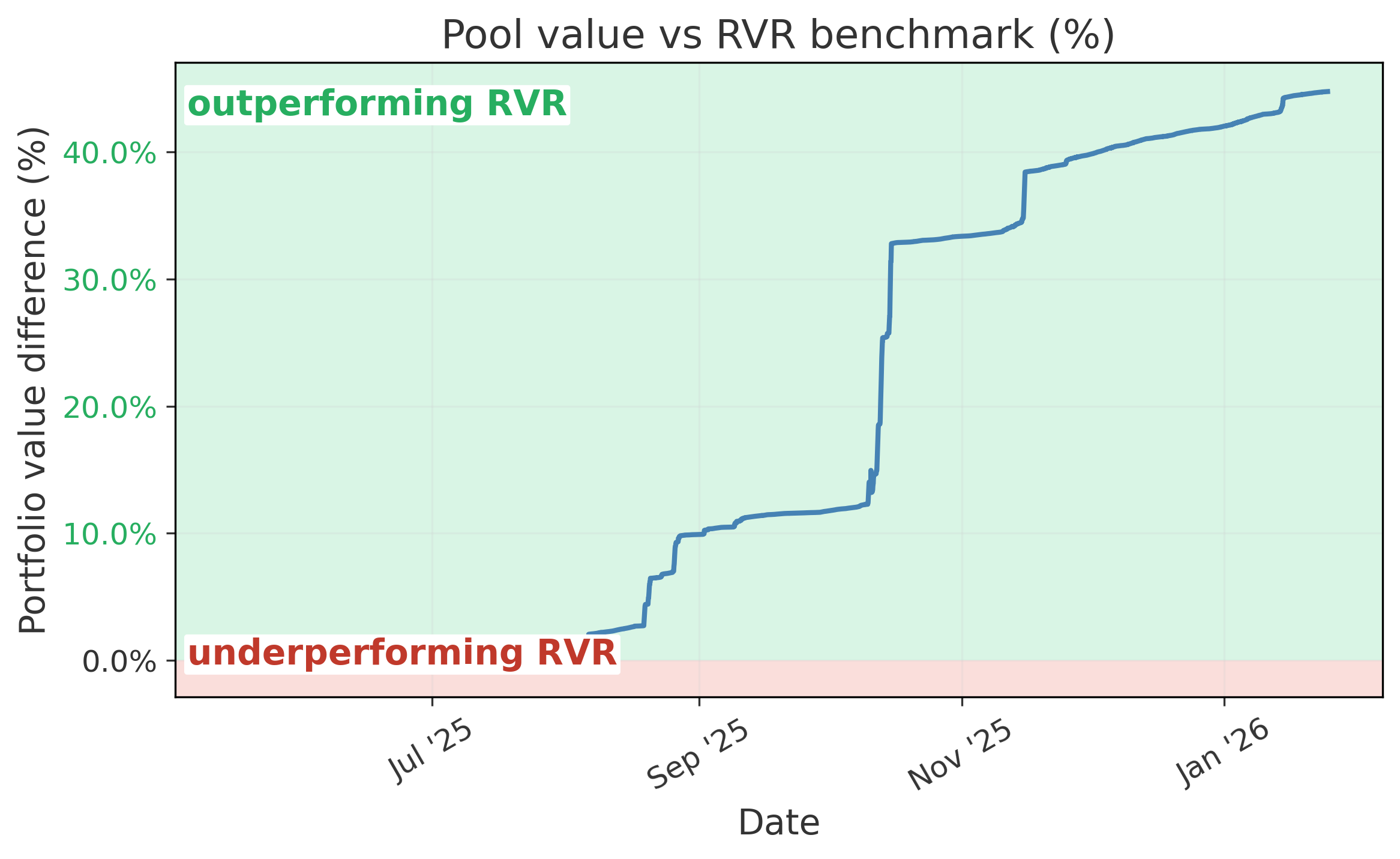}
  \caption{Base L2, same period.}
  \end{subfigure}
  \end{minipage}
  \end{minipage}
\caption{Cumulative portfolio value of each pool relative to LVR and RVR benchmarks. LVR~\cite{lvr} assumes frictionless, continuous rebalancing on a CEX executing the same weight trajectory; RVR~\cite{rvr} adds realistic spreads and commissions. Values above zero indicate the AMM pool outperforming the benchmark over the period.
\textbf{Top row:} Safe Haven (Ethereum mainnet, August 2025--January 2026). \textbf{Bottom row:} Base Macro (Base L2, May 2025--January 2026). Both pools outperform the realistic RVR benchmark, confirming that AMM-based execution is competitive with CEX rebalancing.
The L2 pool additionally outperforms the frictionless LVR benchmark, consistent with the near or below-zero per-trade costs documented in Section~\ref{ssec:block_level}.}
\label{fig:benchmarks}
\end{figure}

Over the period August 2025 to January 2026, the Base Macro exceeded both the LVR benchmark~\cite{lvr} (a hypothetical frictionless CEX rebalancer executing the same weight trajectory) and the RVR benchmark (a realistic CEX rebalancer that pays commissions and crosses spreads).
Figure~\ref{fig:benchmarks} shows cumulative outperformance of approximately 27 percentage points vs.\ LVR and 43 percentage points vs.\ RVR.
This period included episodes of very high volume on Base, which the dynamic-weight strategy was positioned to capture.

These figures reflect the \emph{combined} effect of the pool's dynamic-weight strategy and its execution efficiency; the comparison does not isolate the rebalancing cost component.
The fact that the pool outperforms even the frictionless LVR benchmark indicates that the strategy's signal-driven weight changes more than compensate for arb extraction, and the widening gap suggests that execution costs are not eroding returns.

LVR and RVR attempt to benchmark rebalancing efficiency and so they help answer whether dynamic weight AMMs are good portfolio construction infrastructure. 
Impermanent loss (IL), meanwhile, measures absolute returns against a counterfactual HODL.
Indeed, a fixed-weight (constant-mix) strategy with perfect LVR-style rebalancing can in some market conditions beat HODL.\footnote{Further, is not clear always how to correctly generalise IL when the pool is changing it target weights over time.}
IL-type benchmarks thus mix the choice of strategy (how the weights change, or do not change) with rebalancing efficiency.
To the extent they are useful, they are most relevant to a pool's creator and/or manager in terms of absolute target weight decisions.

\section{Discussion}
\label{sec:discussion}

\paragraph{The sawtooth pattern}
\label{app:sawtooth_schematic}

Over a series of auction cycles, the allocation gap traces a characteristic sawtooth wave (Figure~\ref{fig:auction_schematic}).
Each tooth represents one complete auction cycle: a linear ramp as the gap grows block-by-block, followed by an instantaneous reset when an arbitrageur acts.

The height of each tooth---the gap at which an arbitrage trade occurs---is determined by the leanest arbitrageur's profit threshold.
The shaded triangles above the gas cost line represent the arbitrageurs net profit: the difference between the total gap and the cost of executing the trade.
Compare this schematic with the empirical sawtooth patterns in Figures~\ref{fig:safe_haven_july} and~\ref{fig:safe_haven_jan}: the July data shows tall teeth (large gaps, ${\sim}$6-minute cycles) while January shows compressed teeth (smaller gaps, ${\sim}$90-second cycles), exactly as predicted by the expecting competition between arbitrageurs.

\paragraph{Profitability thresholding as a diagnostic}

The threshold defined in Section~\ref{sec:threshold} serves a dual purpose.
Under the assumption that the pool is the sole source of profit, the threshold defines the minimum extraction needed for a standalone arb to be viable.

In July, this assumption holds cleanly: 100\% of Safe Haven trades have empirical profit above the realistic threshold (Figure~\ref{fig:safe_haven_july}, second panel).
In January, 32\% of trades fall \emph{below} the threshold---viable only because the arbitrageur is running a multi-venue strategy where gas is amortised across legs.
On Base Macro, nearly 100\% of trades fall below the threshold: the pool is rebalanced entirely through incidental routing.

The fraction of trades below the standalone threshold thus provides a simple, observable metric for how much a pool benefits from the routing ecosystem.
As the DEX ecosystem matures~\cite{diamandis, angeris2022optimalrouting}, more rebalancing shifts from targeted extraction to incidental routing, and the pool's rebalancing cost falls without any change to its own fee or design.

\paragraph{Maturity and Efficiency Summary}

New invariants or auction approaches can have a large gap between theoretical perfect efficiency and actual arbitrage discovery and efficiency.
For TFMMs, where in a given block the arbitrage opportunity is presented as if via a completely vanilla G3M, we see that arbitrageurs are consistent and efficient.
Sawtooth patterns in the arbitrage process, most clearly on mainnet, give the clearest picture.
As expected, high frequency, noisier arbitrage is seen when on an L2.

\paragraph{LVR and RVR benchmarks}

LVR is a simple and clean benchmark (do you beat perfection), but its very simplicity is a limitation: as it does not take into account spreads or costs such as fees for centralised rebalancing, it is of perhaps limited use for comparing or analysing the real-world trade offs of different ways to rebalance a portfolio.
RVR was introduced as an improved model for those evaluating execution mechanisms for practical use.
Despite this, on an L2 rebalancing a portfolio as an AMM beats the perfect rebalancing of LVR.

Simulations in~\cite{rvr}, themselves based on empirical study~\cite{canidio2024arbitrageursprofitslvrsandwich}, do appear to have over-estimated retail volume as a ratio of arbitrage volume at the lower pool TVLs studied here.
This may just be a property of price impact and liquidity depth: noise trade assumptions might still be accurate at higher TVLs.
While noise trading was an over-estimate, upgrades and the rich MEV ecosystem have meant that effective gas costs and trade costs have also been an over-estimated for both mainnet and L2s.
With reduced gas costs the threshold required for dynamic weight AMMs to become more efficient than realistic CEX models (RVR) is lowered.
Practically for L2s we have seen this reduction in gas cost be more important than the reduced estimate of noise models.
More accurate noise modelling has also not been explored in either these empirical results or in simulation work.
Finally, note that simulated work and the empirical analysis here also only tests linear interpolation, non-linear methods exist~\cite{willetts2024optimalrebalancingdynamicamms} that can further reduce the barrier for dynamic weight AMMs to beat realistic and perfect CEX based price taker rebalancing.

\newpage
\section{Conclusion}

TFMM pools pay for rebalancing through arbitrage: small, predictable Dutch reverse auctions that run continuously on-chain.
Across two live TFMM/QuantAMM pools observed over time we find consistent auction compression: trades become more frequent, extraction per trade falls, transactions become more multi-venue, and a growing share of trades fall below the standalone profitability threshold.

On Ethereum mainnet (Safe Haven pool), the trade rate increases from 20 in a two hour window to 78, while mean per-trade extraction drops from $\$2.58$ to $\$0.28$, consistent with intensified competition lowering the profit threshold.
On Base (Base Macro pool), extraction is essentially zero (98\% of trades extract $< \$0.01$, many $<\$0$ so in fact they are subsidising the pool) yet the pool still rebalances, indicating a second regime where rebalancing is delivered largely via incidental routing rather than targeted extraction.
These patterns are consistent with external ecosystem maturation
so rebalancing costs can decline over time without changes to pool fee or design, though broader generalisation would require studying more pools, in more market conditions.

\newpage
\bibliography{biblio}

\begin{thebibliography}{16}
\providecommand{\natexlab}[1]{#1}
\providecommand{\url}[1]{\texttt{#1}}
\expandafter\ifx\csname urlstyle\endcsname\relax
  \providecommand{\doi}[1]{doi: #1}\else
  \providecommand{\doi}{doi: \begingroup \urlstyle{rm}\Url}\fi

\bibitem[Willetts and Harrington(2024{\natexlab{a}})]{rvr}
Matthew Willetts and Christian Harrington.
\newblock Rebalancing-versus-rebalancing: Improving the fidelity of loss-versus-rebalancing, 2024{\natexlab{a}}.
\newblock URL \url{https://arxiv.org/abs/2410.23404}.

\bibitem[{IBM}(2024)]{dutchreverse}
{IBM}.
\newblock Dutch reverse auction, 2024.
\newblock URL \url{https://www.ibm.com/docs/en/emptoris-sourcing/10.1.0?topic=rt-dutch-reverse-auction}.
\newblock IBM Emptoris Sourcing documentation.

\bibitem[Willetts and Harrington(2024{\natexlab{b}})]{willetts2024closedform}
Matthew Willetts and Christian Harrington.
\newblock Closed-form solutions for generic n-token amm arbitrage, 2024{\natexlab{b}}.
\newblock URL \url{https://arxiv.org/abs/2402.06731}.

\bibitem[Willetts and Harrington(2024{\natexlab{c}})]{willetts2024optimalrebalancingdynamicamms}
Matthew Willetts and Christian Harrington.
\newblock Optimal rebalancing in dynamic amms, 2024{\natexlab{c}}.
\newblock URL \url{https://arxiv.org/abs/2403.18737}.

\bibitem[Milionis et~al.(2022)Milionis, Moallemi, Roughgarden, and Zhang]{lvr}
Jason Milionis, Ciamac~C. Moallemi, Tim Roughgarden, and Anthony~Lee Zhang.
\newblock Automated market making and loss-versus-rebalancing, 2022.

\bibitem[Milionis et~al.(2023)Milionis, Moallemi, and Roughgarden]{lvr_arb}
Jason Milionis, Ciamac~C. Moallemi, and Tim Roughgarden.
\newblock Automated market making and arbitrage profits in the presence of fees, 2023.
\newblock URL \url{https://arxiv.org/abs/2305.14604}.

\bibitem[Daian et~al.(2020)Daian, Goldfeder, Kell, Li, Zhao, Bentov, Breidenbach, and Juels]{daian2020flashboys}
Philip Daian, Steven Goldfeder, Tyler Kell, Yunqi Li, Xueyuan Zhao, Iddo Bentov, Lorenz Breidenbach, and Ari Juels.
\newblock Flash boys 2.0: Frontrunning in decentralized exchanges, miner extractable value, and consensus instability.
\newblock In \emph{2020 IEEE Symposium on Security and Privacy (SP)}, pages 910--927. IEEE, 2020.

\bibitem[Evans(2019)]{evansG3Ms}
Alex Evans.
\newblock Liquidity provider returns in geometric mean markets, 2019.

\bibitem[Martinelli and Mushegian(2019)]{balancer}
Fernando Martinelli and Nikolai Mushegian.
\newblock Balancer: A non-custodial portfolio manager, liquidity provider, and price sensor., 2019.

\bibitem[Angeris et~al.(2020)Angeris, Evans, and Chitra]{angeris2020does}
Guillermo Angeris, Alex Evans, and Tarun Chitra.
\newblock When does the tail wag the dog? curvature and market making, 2020.

\bibitem[team(2023{\natexlab{a}})]{tfmm}
QuantAMM team.
\newblock Temporal-function market making, 2023{\natexlab{a}}.

\bibitem[team(2023{\natexlab{b}})]{tfmm_litepaper}
QuantAMM team.
\newblock Temporal-function market making litepaper, 2023{\natexlab{b}}.
\newblock URL \url{https://www.quantamm.fi/litepapers}.

\bibitem[{Flashbots}(2021)]{flashbots}
{Flashbots}.
\newblock Flashbots: Frontrunning the mev crisis, 2021.
\newblock URL \url{https://writings.flashbots.net/}.

\bibitem[Diamandis et~al.(2023)Diamandis, Resnick, Angeris, Chitra, Evans, and Boyd]{diamandis}
Theo Diamandis, Max Resnick, Guillermo Angeris, Tarun Chitra, Alex Evans, and Stephen Boyd.
\newblock An efficient algorithm for optimal routing through constant function market makers, 2023.

\bibitem[Angeris et~al.(2022)Angeris, Chitra, Evans, and Boyd]{angeris2022optimalrouting}
Guillermo Angeris, Tarun Chitra, Alex Evans, and Stephen Boyd.
\newblock Optimal routing for constant function market makers, 2022.

\bibitem[Canidio and Fritsch(2024)]{canidio2024arbitrageursprofitslvrsandwich}
Andrea Canidio and Robin Fritsch.
\newblock Arbitrageurs' profits, lvr, and sandwich attacks: batch trading as an amm design response, 2024.
\newblock URL \url{https://arxiv.org/abs/2307.02074}.

\end{thebibliography}

\newpage
\begin{appendices}
\setcounter{figure}{0}
\setcounter{equation}{0}
\setcounter{table}{0}
\renewcommand\thefigure{\thesection.\arabic{figure}}    
\renewcommand\theequation{\thesection.\arabic{equation}}   
\renewcommand\thetable{\thesection.\arabic{table}}   

\section{Transaction structure}
\label{app:tx_structure}

\subsection{Evolving DEX Arb landscape}
Even in July 2025, none of the 22 Safe Haven transactions are naive single-venue swaps.
Every transaction is a multi-hop arb routing through the broader DEX ecosystem.
A typical route sends PAXG to the Safe Haven pool and receives WBTC, sells WBTC for WETH on Uniswap V3, sells WETH for USDC, then buys PAXG with USDC to close the loop---four venues in a single atomic transaction.

No retail or organic swap flow was detected in either window: every transaction originates from professional MEV infrastructure, identified by contract-mediated execution, multi-hop routing through 2+ DEXs, and the absence of direct EOA-initiated single-pool swaps.
This means the pool earns no swap fee revenue from non-arb flow; its returns derive entirely from the value generated by dynamic weight changes, net of arb extraction.

By January, the routing landscape has shifted dramatically.
Uniswap V4 accounts for 37\% of all swap events (up from zero), KyberSwap Elastic has entered the routing table at 11\%, and complex MEV bundles with 4+ swap legs account for 24\% of all trades.
The largest single transaction contains 51 swap events across 5 DEX protocols.

\begin{table}[h!]
\centering
\begin{tabular}{lrrrr}
\toprule
& \multicolumn{2}{c}{\textbf{July 2025}} & \multicolumn{2}{c}{\textbf{January 2026}} \\
\cmidrule(lr){2-3} \cmidrule(lr){4-5}
& Count & \% & Count & \% \\
\midrule
\textbf{Transaction type} & & & & \\
\quad 2-leg arb & 8 & 36 & 23 & 29 \\
\quad 3-leg arb & 13 & 59 & 25 & 32 \\
\quad Complex bundle (4+ swaps) & 1 & 5 & 19 & 24 \\
\quad Other & --- & --- & 12 & 15 \\
\midrule
\textbf{DEX swap events} & & & & \\
\quad UniV3 & 27 & 77 & 137 & 42 \\
\quad UniV4 & 0 & 0 & 119 & 37 \\
\quad Kyber & 0 & 0 & 34 & 11 \\
\quad UniV2 & 8 & 23 & 31 & 10 \\
\midrule
\textbf{All-in cost (mean)} & \multicolumn{2}{c}{\$1.08} & \multicolumn{2}{c}{\$0.20} \\
\quad Base fee share & \multicolumn{2}{c}{82\%} & \multicolumn{2}{c}{39\%} \\
\quad Priority fee share & \multicolumn{2}{c}{10\%} & \multicolumn{2}{c}{59\%} \\
\bottomrule
\end{tabular}

\caption{Transaction complexity, DEX event distribution, and cost decomposition, Safe Haven pool. July costs are dominated by network congestion (base fee at ${\sim}$0.55 gwei); January by builder inclusion competition (priority fee, with base fee down $12\times$ to ${\sim}$0.047 gwei). Some arbitrageurs have off-chain builder arrangements that leave no on-chain cost trace.}
\label{tab:tx_dex_costs}
\end{table}

\newpage
\subsection{Mainnet gas cost decomposition}
\label{app:gas}

The cost structure inverts between observation windows:

\begin{table}[h!]
\centering
\begin{tabular}{lrr}
\toprule
\textbf{Cost component} & \textbf{July 2025} & \textbf{January 2026} \\
\midrule
Base fee (burned) & \$0.88 \;(82\%) & \$0.08 \;(39\%) \\
Priority fee (to builder) & \$0.11 \;(10\%) & \$0.12 \;(59\%) \\
Builder tip (coinbase transfer) & \$0.09 \;(8\%) & \$0.01 \;(2\%) \\
\midrule
Mean all-in cost & \$1.08 & \$0.20 \\
\bottomrule
\end{tabular}
\caption{All-in cost decomposition, Safe Haven pool. July is dominated by the base fee (network congestion at ${\sim}$0.55 gwei mean); January by the priority fee (base fee dropped $12\times$ to ${\sim}$0.047 gwei, making builder inclusion competition the binding constraint). Some arbitrageurs have off-chain arrangements with builders that leave no on-chain cost trace.}
\label{tab:gas_detail}
\end{table}

\subsection{Arber consolidation}
\label{app:consolidation}

A natural concern is that consolidation (from 65 unique arbitrageurs in July to 4 dominant bots by January on Base Macro) signals growing extractive power.
The data points in the opposite direction: consolidation coincides with \emph{lower}, not higher, per-trade extraction.
The dominant bots are sophisticated multi-venue routers.
Their efficiency results in faster rebalancing at lower cost.

\newpage
\section{Arbitrage analysis showing weight changes and cumulative arbitrageur profit}

In addition to the plots in \S\ref{ssec:block_level}, we can show for these same time periods the weight changes and cumulative profits made by arbitrageurs.

\begin{figure}[h]
\centering
\includegraphics[width=\textwidth]{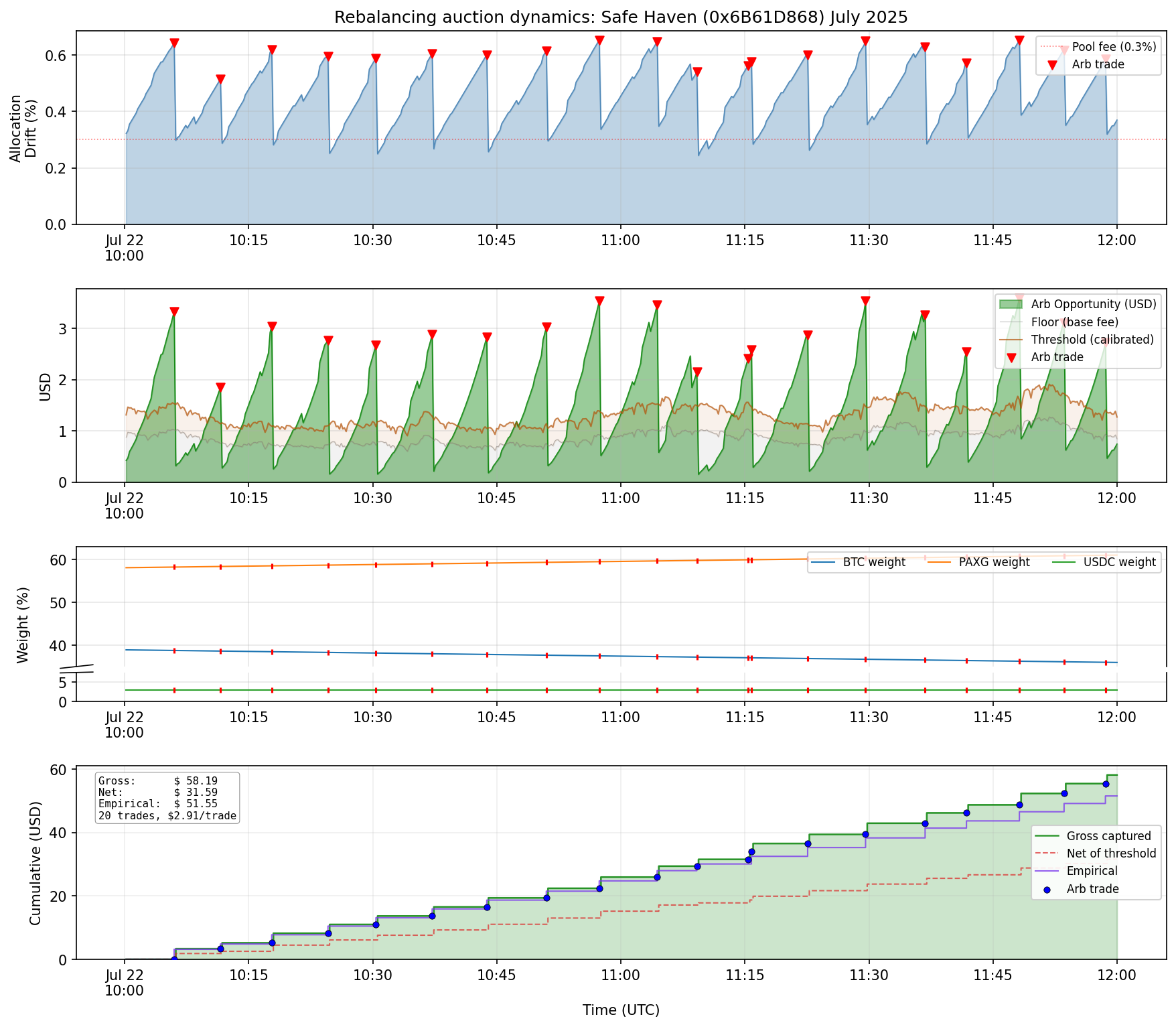}
\caption{Safe Haven pool, July 22 2025 (10:00--12:00 UTC), 606 blocks, 20 arb trades.
\textbf{Top:} Allocation drift sawtooth; dashed line = 0.3\% pool fee.
\textbf{Second:} Per-block arb profit (green) with floor and calibrated thresholds; red triangles = arbitrage trades. All 20 exceed the realistic threshold.
\textbf{Third:} Weight evolution (${\sim}$0.06\% shift over 2 hours).
\textbf{Bottom:} Cumulative extraction---empirical \$51.55 is 88.6\% of theoretical \$58.19.}
\label{fig:4_panel_safe_haven_july}
\end{figure}

\begin{figure}[p]
\centering
\includegraphics[width=\textwidth]{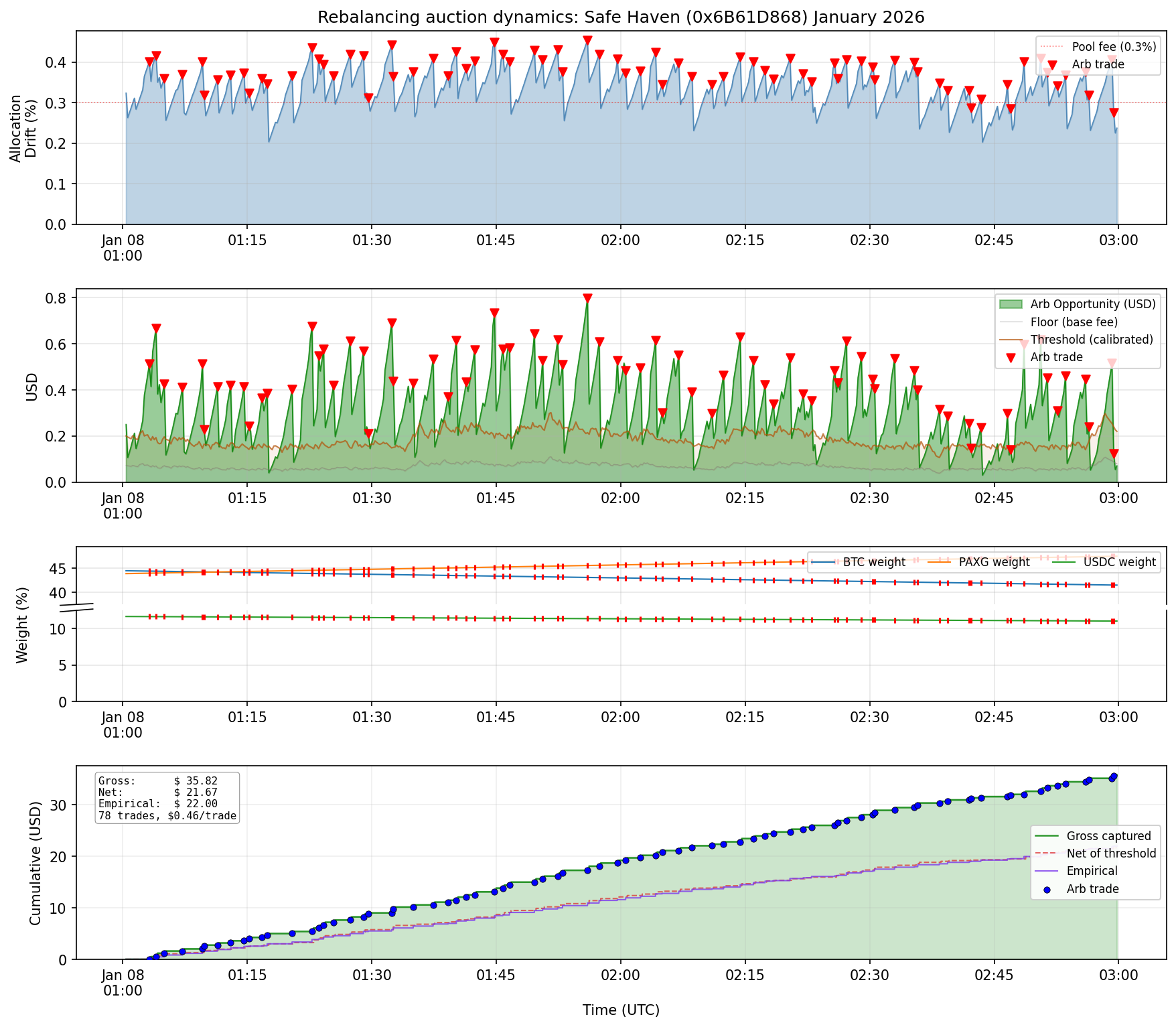}
\caption{Safe Haven pool, January 8 2026 (01:00--03:00 UTC).
}
\label{fig:4_panel_safe_haven_jan}
\end{figure}

\begin{figure}[p]
\centering
\includegraphics[width=\textwidth]{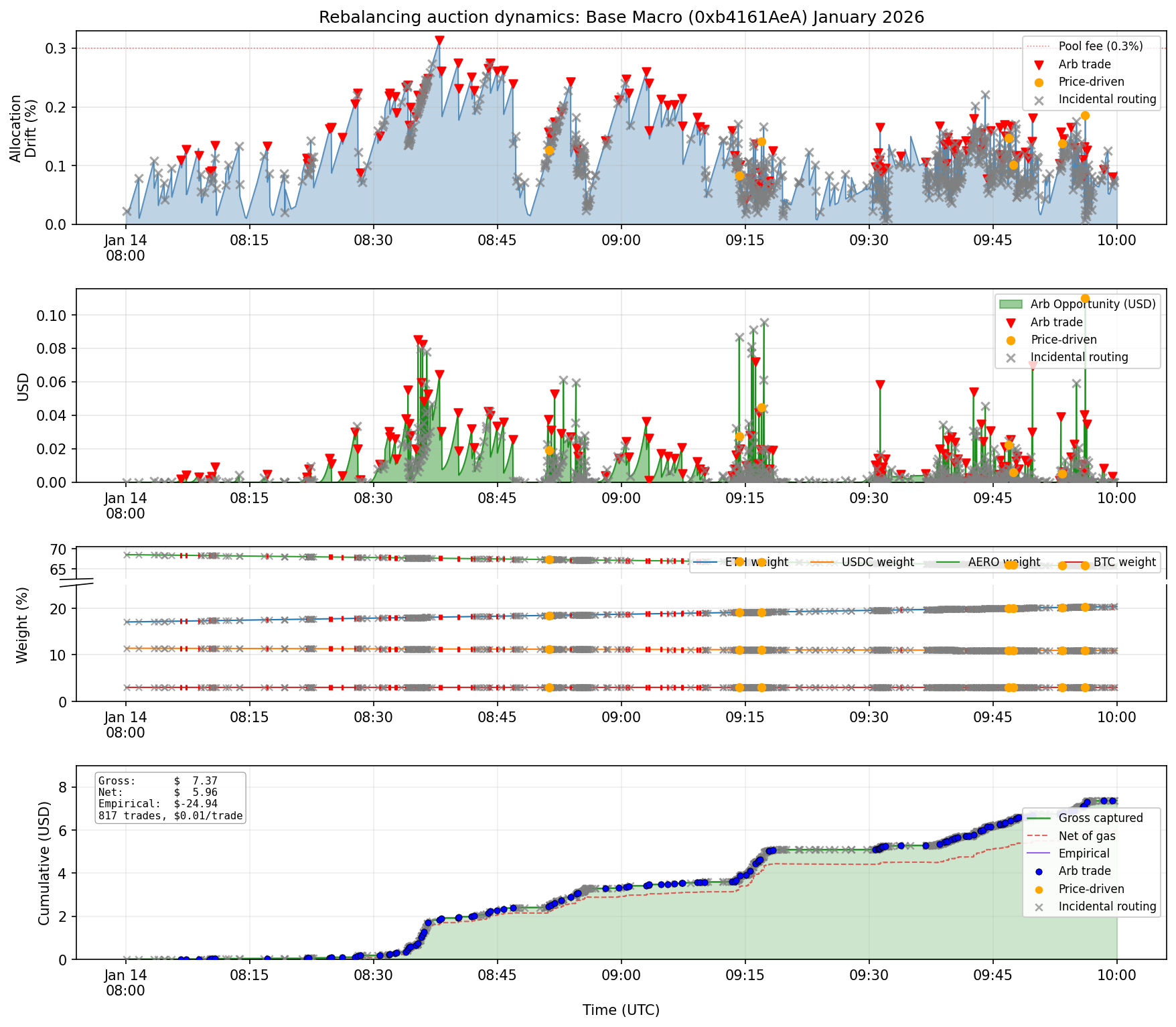}
\caption{Base Macro pool, January 14 2026 (08:00--10:00 UTC).}
\label{fig:4_panel_base_macro_jan}
\end{figure}

\newpage
\section{Practical choices made by Arbitrageurs}
\label{app:undertrading}

As discussed in the main paper, arbitrageurs can get most of the profit of an arb opportunity while under-sizing their trade.
Comparing empirical trades to the theoretical optimum reveals that arbitrageurs consistently under-trade, executing 54--71\% of the optimal trade size while capturing 81--99\% of the available profit.
The return-to-trade function is $\sim$ flat in the region of the optimum.
So this is rational: the last 30\% of the optimal trade adds little marginal profit but requires that additional capital.

Further, we also observe that arbitrageurs use a small set of pre-computed trade sizes rather than solving the optimum each block.
The PAXG buy amount repeats across trades (e.g. multiple trades use the same trade amounts of 0.1307, 0.1283, 0.1017 PAXG), suggesting a strategy of simulating a number of discrete trade sizes that can capture most of the profit with minimal computational overhead.

\section{Finite-size effects on boundary targeting}
\label{app:finite_size}

\begin{figure}[h!]
\centering
\scalebox{0.8}{\begin{tikzpicture}[
  every node/.style={font=\sffamily},
  >={Stealth[length=5pt]}
]

\node[font=\sffamily\bfseries\Large, anchor=west] at (0, 14.5)
  {Why Optimal Arbs Target the Boundary};
\node[font=\sffamily\small, text=axgray, anchor=west, text width=14.5cm] at (0, 13.3)
  {An arb's profit is maximised when the post-trade $m_u$ places $m_p$ at the
   no-arb boundary --- not when $m_u = m_p$.  Trading past that point costs
   more in fees than it earns.};

\begin{scope}[shift={(0.5, 6.5)}]
  \node[font=\sffamily\bfseries, anchor=west] at (0, 6.2)
    {(a) Arb profit vs.\ trade size};

  \draw[->, thick, axgray] (0,0) -- (12.5,0)
    node[below, font=\sffamily\small, xshift=-5pt] {trade size};
  \draw[->, thick, axgray] (0,0) -- (0,5.6)
    node[rotate=90, anchor=south, font=\sffamily\small, pos=0.5] {profit};

  \draw[axgray, thin, dashed] (0,0) -- (12,0);

  \draw[arbprofit, very thick] plot[smooth, tension=0.6] coordinates {
    (0, 0) (1.5, 1.6) (3.0, 3.0) (4.5, 4.1) (5.0, 4.25)
    (5.5, 4.15) (6.5, 3.7) (8.0, 2.6) (9.5, 1.2) (10.5, 0) (11.5, -0.8)
  };

  \fill[arbprofit, opacity=0.06] plot[smooth, tension=0.6] coordinates {
    (0, 0) (1.5, 1.6) (3.0, 3.0) (4.5, 4.1) (5.0, 4.25)
    (5.5, 4.15) (6.5, 3.7) (8.0, 2.6) (9.5, 1.2) (10.5, 0)
  } -- (0, 0) -- cycle;

  \draw[noarb, thick, dashed] (5.5, -0.5) -- (5.5, 5.0);

  \fill[arbprofit] (5.0, 4.25) circle (5pt);
  \node[arbprofit, font=\sffamily\footnotesize\bfseries, anchor=south east,
        xshift=-3pt, yshift=2pt] at (5.0, 4.25) {optimum};

  \draw[axgray, <->, line width=0.8pt] (5.05, 4.55) -- (5.45, 4.55);
  \node[axgray, font=\sffamily\tiny, above=0pt] at (5.25, 4.55) {\textit{price impact}};

  \draw[poolprice, thick, dashed] (10.5, -0.5) -- (10.5, 0);
  \fill[poolprice] (10.5, 0) circle (5pt);
  \node[poolprice, font=\sffamily\footnotesize\bfseries, anchor=south west]
    at (10.6, 0.3) {``naive'' ($m_u\!=\!m_p$)};

  \draw[axgray, thin, densely dashed] (5.0, 4.25) -- (10.5, 4.25);
  \draw[decorate, decoration={brace, amplitude=5pt}, thick, axgray]
    (10.7, 4.25) -- (10.7, 0.15);
  \node[axgray, font=\sffamily\footnotesize, anchor=west, text width=2cm] at (11.1, 2.2)
    {profit lost\\by overtrading};

  \draw[gascost, thick, ->] (2.5, 5.4) -- (4.8, 5.4);
  \node[gascost, font=\sffamily\scriptsize, above=1pt] at (3.65, 5.4) {marg.\ revenue $>$ fee};
  \draw[arbprofit, thick, ->] (8.5, 5.4) -- (5.7, 5.4);
  \node[arbprofit, font=\sffamily\scriptsize, above=1pt] at (7.1, 5.4) {fee $>$ marg.\ revenue};

  \node[axgray, font=\sffamily\scriptsize, below=6pt] at (0.3, 0) {no trade};
  \node[noarb, font=\sffamily\scriptsize, fill=white, inner sep=1pt, below=6pt] at (5.5, 0) {boundary};

\end{scope}

\begin{scope}[shift={(0.5, -1.0)}]
  \node[font=\sffamily\bfseries, anchor=west] at (0, 6.2)
    {(b) Post-trade pool state: optimal vs.\ naive};

  \begin{scope}[yshift=4.4cm]
    \node[font=\sffamily\bfseries\small, anchor=east] at (1.3, 0) {Before};
    \draw[axgray, thin] (1.5, 0) -- (12.5, 0);
    \fill[bandfill, rounded corners=2pt] (7.5, -0.3) rectangle (11.5, 0.3);
    \draw[noarb, thick, rounded corners=2pt] (7.5, -0.3) rectangle (11.5, 0.3);
    \fill[poolprice] (9.5, 0) circle (3pt);
    \node[poolprice, font=\sffamily\scriptsize, above=3pt] at (9.5, 0.3) {$m_u$};
    \fill[marketprice] (5.0, 0) circle (3pt);
    \node[marketprice, font=\sffamily\scriptsize, above=3pt] at (5.0, 0.3) {$m_p$};
    \fill[dangerfill, opacity=0.4] (5.0, -0.3) rectangle (7.5, 0.3);
    \fill[marketprice] (5.0, 0) circle (3pt);
    \draw[arbprofit, very thick, <->] (5.0, -0.6) -- (7.5, -0.6);
    \node[arbprofit, font=\sffamily\scriptsize, below=1pt] at (6.25, -0.6) {full gap};
  \end{scope}

  \begin{scope}[yshift=2.6cm]
    \node[font=\sffamily\bfseries\small, anchor=east, noarb] at (1.3, 0) {Optimal};
    \draw[axgray, thin] (1.5, 0) -- (9.0, 0);
    \fill[bandfill, rounded corners=2pt] (5.0, -0.3) rectangle (8.2, 0.3);
    \draw[noarb, very thick, rounded corners=2pt] (5.0, -0.3) rectangle (8.2, 0.3);
    \fill[poolprice] (6.6, 0) circle (3pt);
    \node[poolprice, font=\sffamily\scriptsize, above=3pt] at (6.6, 0.3) {$m_u'$};
    \fill[marketprice] (5.0, 0) circle (3pt);
    \node[marketprice, font=\sffamily\scriptsize, above=3pt] at (5.0, 0.3) {$m_p$};
    \draw[noarb, very thick, <->] (5.0, -0.6) -- (6.6, -0.6);
    \node[noarb, font=\sffamily\footnotesize\bfseries, below=1pt] at (5.8, -0.6) {$\approx$ fee};
    \node[noarb, font=\sffamily\footnotesize, anchor=west] at (9.5, 0)
      {$m_p \approx \gamma\, m_u'$ \textbf{(max profit)}};
  \end{scope}

  \begin{scope}[yshift=0.8cm]
    \node[font=\sffamily\bfseries\small, anchor=east, poolprice] at (1.3, 0) {Naive};
    \draw[axgray, thin] (1.5, 0) -- (9.0, 0);
    \fill[bandfill, rounded corners=2pt] (3.4, -0.3) rectangle (6.6, 0.3);
    \draw[noarb, thick, rounded corners=2pt] (3.4, -0.3) rectangle (6.6, 0.3);
    \fill[poolprice] (5.0, 0) circle (3pt);
    \node[poolprice, font=\sffamily\scriptsize, above=3pt] at (5.0, 0.3) {$m_u''\!=\!m_p$};
    \fill[marketprice] (5.0, 0) circle (3pt);
    \draw[poolprice, thick] (5.0, -0.5) -- (5.0, -0.7);
    \node[poolprice, font=\sffamily\footnotesize, below=1pt] at (5.0, -0.7) {gap $= 0$};
    \node[poolprice, font=\sffamily\footnotesize, anchor=west] at (9.5, 0)
      {$m_u''\!=\!m_p$ (overtraded, less profit)};
  \end{scope}

  \draw[noarb, very thick, ->] (1.8, 3.0) -- (1.8, 1.2);
  \node[noarb, font=\sffamily\scriptsize, rotate=90, anchor=south] at (1.5, 2.1) {more profit};

\end{scope}

\begin{scope}[shift={(0, -2.7)}]
  \draw[rounded corners=6pt, thick, axgray, fill=bgfill]
    (0.3, -1.3) rectangle (14.5, 1.3);
  \node[font=\sffamily\small, text width=13.5cm, align=left] at (7.4, 0) {
    \textbf{Intuition:} Each marginal unit of trade yields less revenue
    (the AMM's marginal price moves against you) but costs the same fee.
    The optimum is where marginal revenue $=$ fee --- approximately when
    $m_p$ reaches the no-arb boundary $\gamma\, m_u'$.
    For finite-sized trades, cumulative price impact shifts the true optimum
    slightly \emph{inside} the boundary; this slippage grows linearly with
    trade size (${\sim}\,0.5\,$bp per 1\% of TVL traded).
    In the limit of small trades the optimum \emph{is} the boundary.
  };
\end{scope}

\end{tikzpicture}}
\caption{Why optimal arbitrage trades target the no-arb boundary.
(a)~Arb profit peaks near the trade size that places $m_p$ at the boundary $\gamma m_u'$, then declines; a ``naive'' arb that trades to $m_u = m_p$ earns zero.
The profit function is extremely flat near the optimum, explaining why arbitrageurs can under-trade substantially (54--71\% of optimal size) while capturing most of the profit (81--99\%).
(b)~Number-line comparison of pre-trade, optimal post-trade, and naive post-trade states.
The optimal arb leaves a residual gap $\approx$ fee; the naive arb eliminates the gap entirely at the cost of zero profit.}
\label{fig:boundary_targeting}
\end{figure}

One might think that the optimal arb trade results in the pool quoting market prices post trade.
After all, that is when pool value is minimised.
For pool with zero fees, that is true, but when fees are charged an optimal arbtrade lands the pool's post-trade quoted prices near but not exactly at the boundary of the no trade region.
Figure~\ref{fig:boundary_targeting} describes why pool's are taken (near the) the boundary at all.

But why not exactly the boundary?
In fact, for infinitesimal trades the optimal trade does take post-trade quoted prices exactly to the edge.
For trades of finite size relative to pool TVL, cumulative price impact shifts the profit-maximising post-trade price ratio slightly \emph{inside} the no-arb region.
 
\begin{figure}[p]
\centering
\begin{subfigure}[t]{\textwidth}
\includegraphics[width=\textwidth]{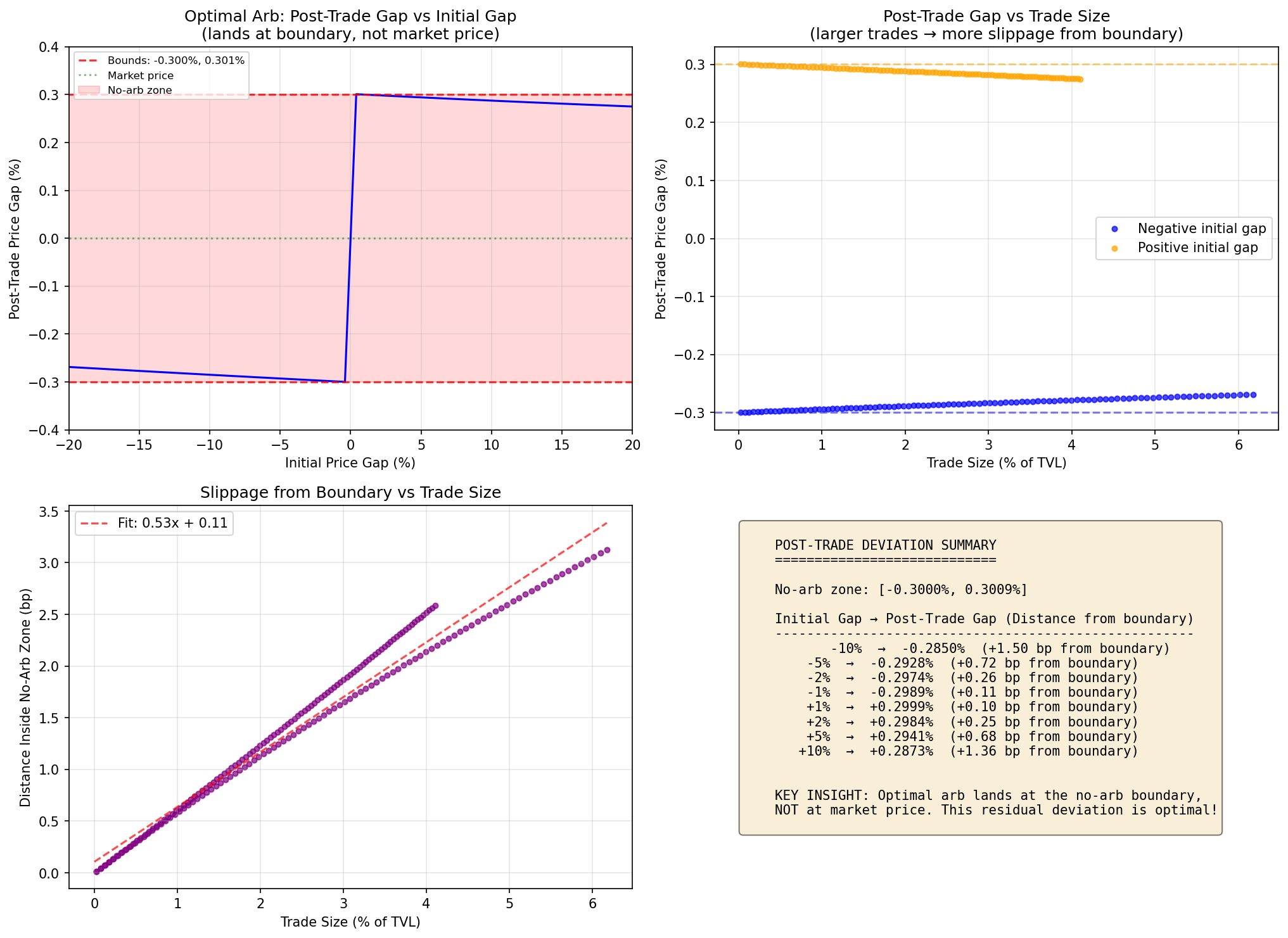}
\caption{Post-trade gap analysis. \emph{Top left:} post-trade gap vs.\ initial gap lands near, not at, the boundary (red dashed lines). \emph{Top right:} larger trades (larger initial gaps) push the post-trade ratio further inside the band. \emph{Bottom left:} slippage from boundary grows linearly with trade size at ${\sim}\,0.5\,$bp per 1\% TVL.}
\end{subfigure}
\hfill
\begin{subfigure}[t]{\textwidth}
\includegraphics[width=\textwidth]{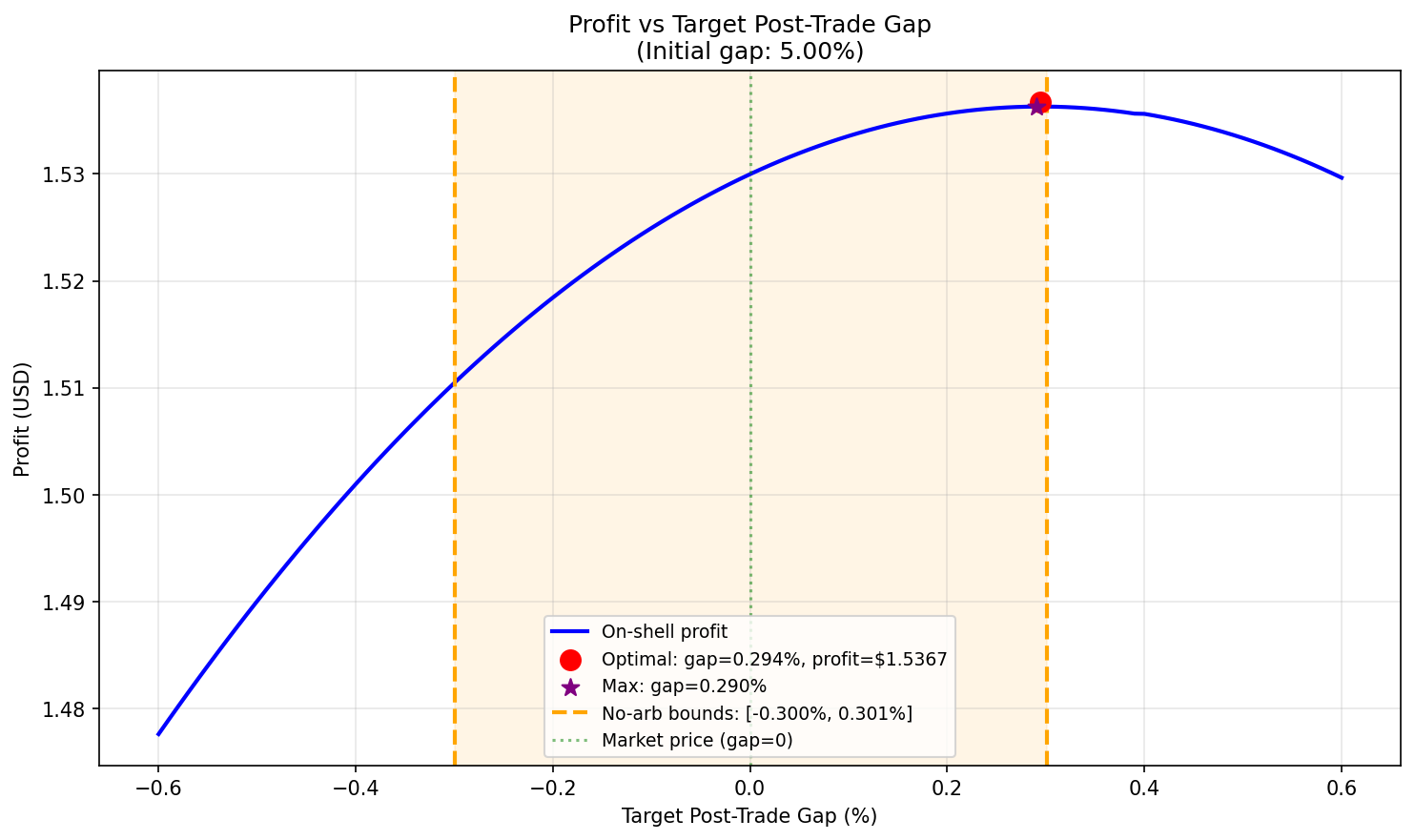}
\caption{Profit vs.\ target post-trade gap for a 5\% initial deviation. The profit maximum (red dot) falls just inside the no-arb boundary (orange dashed), not at market price (green dotted, gap $= 0$). The curve is extremely flat near the maximum, explaining why arbitrageurs can under-trade substantially without sacrificing much profit.}
\end{subfigure}
\caption{Finite-size effects on optimal arbitrage in a two-token G3M pool (fee $= 0.3\%$, equal weights). The optimal post-trade price ratio lies slightly inside the no-arb boundary, with the deviation growing linearly in trade size. In the limit of small trades, the optimum converges to the boundary exactly.}
\label{fig:finite_size}
\end{figure}

Figure~\ref{fig:finite_size} illustrates this for a two-token equal-weight pool with a 0.3\% fee ($\gamma = 0.997$).
Panel~(a) shows the post-trade price gap as a function of the initial gap: for any starting deviation outside the band, the optimal trade brings the pool to a post-trade gap that sits just inside $[\gamma, \gamma^{-1}]$, not exactly at its edge.
Panel~(b) plots the distance from the boundary as a function of trade size (as a fraction of TVL).
The relationship is approximately linear at ${\sim}\,0.5$ basis points per 1\% of TVL traded.
Panel~(c) shows why this is optimal: the on-shell profit curve as a function of target post-trade gap peaks just inside the boundary (red dot), not at the boundary itself (dashed orange lines) and not at market price (green dotted line at gap $= 0$).
Trading past the optimum toward the boundary would cost more in price impact than it earns from the smaller residual gap.

For the pools studied in this paper the effect is small---typical arb trades are ${\ll}\,1\%$ of TVL, so the optimum is within ${\sim}\,0.1$ bp of the boundary---but it is a systematic bias, not noise.
In the multivariate case the picture is richer: only the pair actively traded is driven to the boundary; other pairwise ratios may remain well inside their bands.

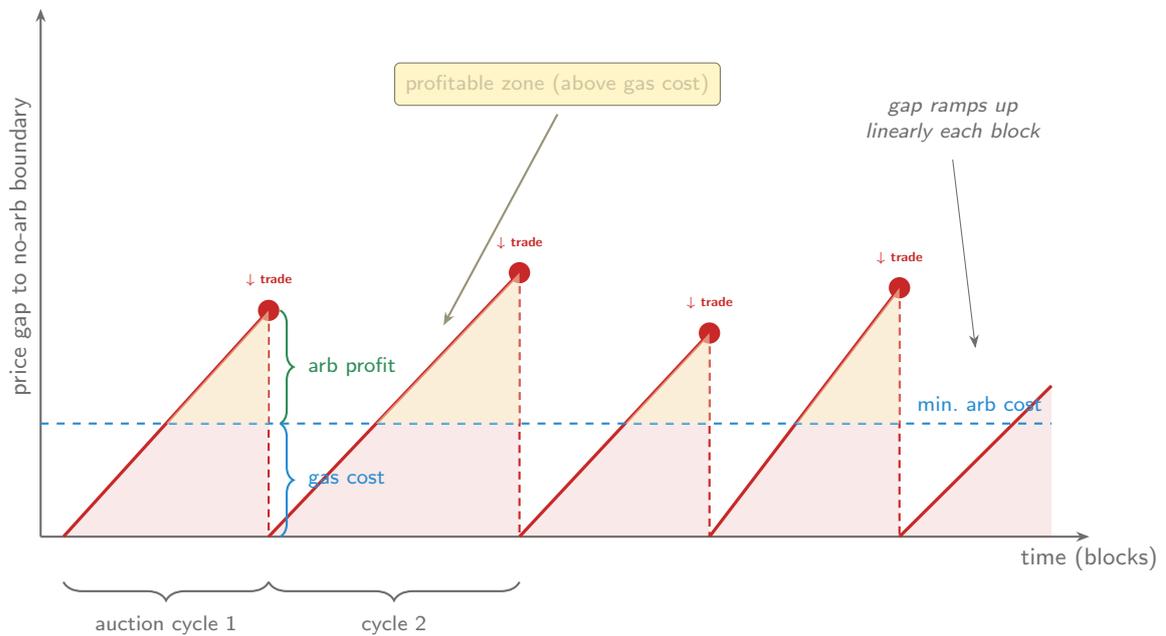
\begin{figure}[p]
\centering
\begin{tikzpicture}[
  every node/.style={font=\sffamily},
  >={Stealth[length=5pt]}
]

\node[font=\sffamily\bfseries\Large, anchor=west] at (0, 9.5)
  {The Sawtooth Pattern --- Continuous View};
\node[font=\sffamily\small, text=axgray, anchor=west, text width=14.5cm] at (0, 8.3)
  {Over many blocks the ``reverse Dutch auction'' produces a characteristic
   sawtooth wave in the price gap.  Each tooth = one complete auction cycle.
   Faster weight change $\Rightarrow$ steeper ramp $\Rightarrow$ more frequent arb trades.};

\def\ox{1.2}
\def\oy{0.5}

\draw[->, thick, axgray] (\ox,\oy) -- (15,\oy)
  node[below, font=\sffamily\small] {time (blocks)};
\draw[->, thick, axgray] (\ox,\oy) -- (\ox,7.5)
  node[rotate=90, anchor=south, font=\sffamily\small, pos=0.55]
  {price gap to no-arb boundary};

\fill[arbprofit, opacity=0.10] (1.5, \oy) -- (4.2, 3.5) -- (4.2, \oy) -- cycle;
\draw[arbprofit, very thick] (1.5, \oy) -- (4.2, 3.5);
\draw[arbprofit, thick, densely dashed] (4.2, 3.5) -- (4.2, \oy);
\fill[arbprofit] (4.2, 3.5) circle (4pt);

\fill[arbprofit, opacity=0.10] (4.2, \oy) -- (7.5, 4.0) -- (7.5, \oy) -- cycle;
\draw[arbprofit, very thick] (4.2, \oy) -- (7.5, 4.0);
\draw[arbprofit, thick, densely dashed] (7.5, 4.0) -- (7.5, \oy);
\fill[arbprofit] (7.5, 4.0) circle (4pt);

\fill[arbprofit, opacity=0.10] (7.5, \oy) -- (10.0, 3.2) -- (10.0, \oy) -- cycle;
\draw[arbprofit, very thick] (7.5, \oy) -- (10.0, 3.2);
\draw[arbprofit, thick, densely dashed] (10.0, 3.2) -- (10.0, \oy);
\fill[arbprofit] (10.0, 3.2) circle (4pt);

\fill[arbprofit, opacity=0.10] (10.0, \oy) -- (12.5, 3.8) -- (12.5, \oy) -- cycle;
\draw[arbprofit, very thick] (10.0, \oy) -- (12.5, 3.8);
\draw[arbprofit, thick, densely dashed] (12.5, 3.8) -- (12.5, \oy);
\fill[arbprofit] (12.5, 3.8) circle (4pt);

\draw[arbprofit, very thick] (12.5, \oy) -- (14.5, 2.5);
\fill[arbprofit, opacity=0.10] (12.5, \oy) -- (14.5, 2.5) -- (14.5, \oy) -- cycle;

\pgfmathsetmacro{\gasline}{\oy + 1.5}
\draw[gascost, thick, dashed] (\ox, \gasline) -- (14.5, \gasline);
\node[gascost, font=\sffamily\footnotesize, anchor=south east]
  at (14.5, \gasline+0.05) {min.\ arb cost};

\fill[strikefill, opacity=0.5] (2.85, \gasline) -- (4.2, 3.5) -- (4.2, \gasline) -- cycle;
\fill[strikefill, opacity=0.5] (5.61, \gasline) -- (7.5, 4.0) -- (7.5, \gasline) -- cycle;
\fill[strikefill, opacity=0.5] (8.89, \gasline) -- (10.0, 3.2) -- (10.0, \gasline) -- cycle;
\fill[strikefill, opacity=0.5] (11.14, \gasline) -- (12.5, 3.8) -- (12.5, \gasline) -- cycle;

\draw[decorate, decoration={brace, amplitude=7pt, mirror}, thick, axgray]
  (1.5, -0.1) -- (4.2, -0.1);
\node[axgray, font=\sffamily\footnotesize, below=9pt, text width=2.5cm, align=center]
  at (2.85, -0.1) {auction cycle 1};

\draw[decorate, decoration={brace, amplitude=7pt, mirror}, thick, axgray]
  (4.2, -0.1) -- (7.5, -0.1);
\node[axgray, font=\sffamily\footnotesize, below=9pt, text width=2.5cm, align=center]
  at (5.85, -0.1) {cycle 2};

\draw[decorate, decoration={brace, amplitude=5pt}, thick, noarb]
  (4.35, 3.5) -- (4.35, \gasline);
\node[noarb, font=\sffamily\footnotesize, right=7pt] at (4.35, {(\gasline+3.5)/2})
  {arb profit};

\draw[decorate, decoration={brace, amplitude=5pt}, thick, gascost]
  (4.35, \gasline) -- (4.35, \oy);
\node[gascost, font=\sffamily\footnotesize, right=7pt] at (4.35, {(\oy+\gasline)/2})
  {gas cost};

\node[arbprofit, font=\sffamily\tiny\bfseries] at (4.2, 3.9) {$\downarrow$ trade};
\node[arbprofit, font=\sffamily\tiny\bfseries] at (7.5, 4.4) {$\downarrow$ trade};
\node[arbprofit, font=\sffamily\tiny\bfseries] at (10.0, 3.6) {$\downarrow$ trade};
\node[arbprofit, font=\sffamily\tiny\bfseries] at (12.5, 4.2) {$\downarrow$ trade};

\node[strikefill!80!black, font=\sffamily\footnotesize, fill=strikefill,
      rounded corners=2pt, inner sep=4pt, draw=strikefill!50!black]
  at (8, 6.5) {profitable zone (above gas cost)};
\draw[->, strikefill!60!black, thick] (8, 6.1) -- (6.5, 3.3);

\node[axgray, font=\sffamily\footnotesize\itshape, text width=3.5cm, align=center]
  at (13.2, 6) {gap ramps up\\linearly each block};
\draw[axgray, thin, ->] (13.2, 5.5) -- (13.5, 3.0);

\end{tikzpicture}
\caption{The sawtooth pattern in continuous view.
The price gap to the no-arb boundary ramps up linearly each block, then drops to zero when an arbitrageur acts (dashed vertical lines).
The golden triangles above the gas cost line show the profitable zone.
Right-side braces on the first tooth decompose each trade into arb profit (green) and gas cost (blue).
Faster weight changes produce steeper ramps and more frequent trades.}
\label{fig:sawtooth_schematic}
\end{figure}

\begin{figure}[p]
\centering
\begin{tikzpicture}[
  every node/.style={font=\sffamily},
  >={Stealth[length=5pt]}
]

\node[font=\sffamily\bfseries\Large, anchor=west] at (0, 10.5)
  {Weight Interpolation Reduces Rebalancing Cost};
\node[font=\sffamily\small, text=axgray, anchor=west, text width=14.5cm] at (0, 9.3)
  {Splitting a weight update into $N$ steps reduces total arb cost, because
   arb profit scales \emph{super-linearly} ($\sim \Delta w^2$) with price gap.};

\begin{scope}[shift={(0.5, 3.0)}]
  \node[font=\sffamily\bfseries, anchor=west] at (0, 5.2)
    {(a) $N$-step total cost tiles inside 1-step cost};

  \fill[interpA, opacity=0.10] (1.0, 0.3) rectangle (5.0, 4.3);
  \draw[interpA, very thick] (1.0, 0.3) rectangle (5.0, 4.3);

  \foreach \i in {0,...,4} {
    \pgfmathsetmacro{\xl}{1.0 + \i*0.8}
    \pgfmathsetmacro{\xr}{\xl + 0.8}
    \fill[interpB, opacity=0.15] (\xl, 0.3) rectangle (\xr, 1.1);
    \draw[interpB, thick, opacity=0.5] (\xl, 0.3) rectangle (\xr, 1.1);
  }

  \draw[axgray, thin, densely dashed] (1.0, 1.1) -- (5.0, 1.1);

  \draw[interpA, thick, <->] (5.3, 0.3) -- (5.3, 4.3);
  \node[interpA, font=\sffamily\footnotesize, right=2pt] at (5.3, 2.3) {$\Delta w$};

  \draw[interpB, thick, <->] (6.0, 0.3) -- (6.0, 1.1);
  \node[interpB, font=\sffamily\footnotesize, right=2pt] at (6.0, 0.7)
    {$\frac{\Delta w}{N}$};

  \node[interpA, font=\sffamily\bfseries] at (3.0, 3.2) {cost saved};
  \node[interpA, font=\sffamily\small] at (3.0, 2.5)
    {$\left(1 - \tfrac{1}{N}\right)\!(\Delta w)^2$};

  \node[interpB, font=\sffamily\footnotesize\bfseries] at (3.0, 0.7)
    {$N$-step total: $\frac{(\Delta w)^2}{N}$};

  \node[interpA, font=\sffamily\footnotesize] at (3.0, 4.6)
    {1-step cost: $(\Delta w)^2$};
\end{scope}

\begin{scope}[shift={(8.5, 3.0)}]
  \node[font=\sffamily\bfseries, anchor=west] at (0, 5.2)
    {(b) Why: cost is quadratic in gap};

  \draw[->, thick, axgray] (0, 0) -- (6.0, 0)
    node[right, font=\sffamily\footnotesize] {gap};
  \draw[->, thick, axgray] (0, 0) -- (0, 4.8)
    node[rotate=90, anchor=south, font=\sffamily\footnotesize, pos=0.5]
    {arb cost};

  \draw[axgray, thin, densely dashed] (0, 0) -- (5.0, 4.0);
  \node[axgray, font=\sffamily\tiny, anchor=south east] at (5.0, 4.05) {linear};

  \draw[very thick] plot[smooth, domain=0:5.15, samples=50]
    (\x, {0.16*\x*\x});
  \node[font=\sffamily\footnotesize, anchor=south west] at (5.1, 4.0) {$g^2$};

  \fill[interpA] (5.0, 4.0) circle (4pt);
  \node[interpA, font=\sffamily\footnotesize, below=2pt] at (5.0, 0) {$\Delta w$};
  \draw[interpA, thin, dashed] (5.0, 0) -- (5.0, 4.0);

  \fill[interpB] (1.0, 0.16) circle (4pt);
  \node[interpB, font=\sffamily\footnotesize, below=2pt] at (1.0, 0)
    {$\frac{\Delta w}{N}$};
  \draw[interpB, thin, dashed] (1.0, 0) -- (1.0, 0.8);
  \fill[axgray] (1.0, 0.8) circle (2pt);

  \node[axgray, font=\sffamily\footnotesize, text width=3cm, align=left,
        anchor=north west] at (0.15, 4.4)
    {Quadratic is \emph{flat} at origin:
     small $\Delta w$ costs almost nothing.\\[3pt]
     Linear cost $\Rightarrow$ no benefit from splitting.};

\end{scope}

\begin{scope}[shift={(0.3, -0.7)}]
  \draw[rounded corners=6pt, thick, axgray, fill=bgfill]
    (0, 0) rectangle (14.5, 3.2);

  \node[font=\sffamily\small, text width=14cm, align=left, anchor=north west]
    at (0.3, 3.0) {
    \textbf{Key insight:} For a G3M pool, rebalancing preserves value to
    first order: $V' = V + \mathcal{O}(\delta w^2)$.
    Cost is purely second-order (quadratic, not linear), so splitting
    $\Delta w$ into $N$ steps of $\Delta w / N$ gives:
  };

  \node[font=\normalsize, text=axgray] at (7.25, 1.2) {
    $\displaystyle
    N \times \left(\frac{\Delta w}{N}\right)^{\!2}
    \;=\; \frac{(\Delta w)^2}{N}
    \;\xrightarrow{N \to \infty}\; 0
    $
  };
\end{scope}

\end{tikzpicture}
\caption{Weight interpolation reduces rebalancing cost.
(a)~Splitting $\Delta w$ into $N$ steps: the $N$ small-step costs (blue strip) tile into the bottom of the single-step cost (red square), with savings $= (1 - 1/N)(\Delta w)^2$.
(b)~The reduction works because arb cost scales as $g^2$ (solid curve), which is flat near the origin, unlike a linear cost (dashed).
In the limit, total cost $(\Delta w)^2 / N \to 0$.}
\label{fig:interpolation_cost}
\end{figure}

\newpage
\section{Complete rebalancing lifecycle}
\label{app:lifecycle}

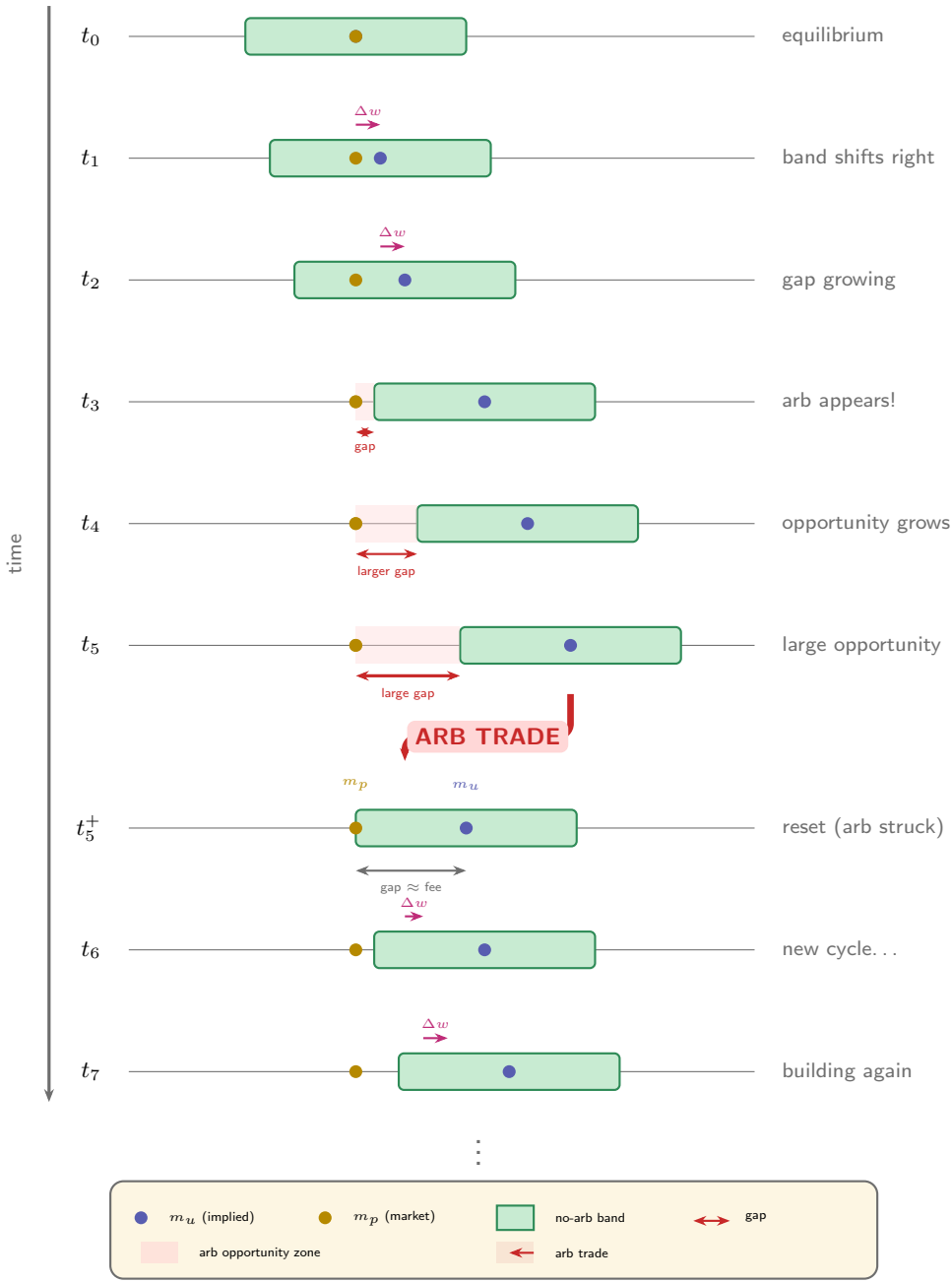
\begin{figure}[p]
\centering
\begin{tikzpicture}[
  every node/.style={font=\sffamily},
  >={Stealth[length=5pt]},
  xscale=0.82, yscale=0.82
]

\node[font=\sffamily\bfseries\Large, anchor=west] at (0, 22.5)
  {Complete Rebalancing Lifecycle --- Detailed View};
\node[font=\sffamily\small, text=axgray, anchor=west, text width=14.5cm] at (0, 21.3)
  {Combining the number-line view of no-arb bands with the auction timeline.
   Watch how the band, implied price, and arb opportunity evolve block-by-block.};

\newcommand{\staterow}[8]{%
  \begin{scope}[yshift=#1 cm]
    \node[font=\sffamily\bfseries\small, anchor=east] at (1.5, 0) {#6};
    \draw[axgray, thin] (1.8, 0) -- (12.0, 0);
    \fill[bandfill, rounded corners=2pt] (#2,-0.3) rectangle (#3,0.3);
    \draw[noarb, thick, rounded corners=2pt] (#2,-0.3) rectangle (#3,0.3);
    \fill[poolprice] (#4, 0) circle (3pt);
    \fill[marketprice] (#5, 0) circle (3pt);
    \node[axgray, font=\sffamily\footnotesize, anchor=west] at (12.3, 0) {#7};
  \end{scope}
}

\staterow{20}{3.7}{7.3}{5.5}{5.5}{$t_0$}{equilibrium}{0}

\staterow{18}{4.1}{7.7}{5.9}{5.5}{$t_1$}{band shifts right}{0}
\draw[weightchange, thick, ->] (5.5, 18.55) -- (5.9, 18.55)
  node[above, midway, font=\sffamily\tiny] {$\Delta w$};

\staterow{16}{4.5}{8.1}{6.3}{5.5}{$t_2$}{gap growing}{0}
\draw[weightchange, thick, ->] (5.9, 16.55) -- (6.3, 16.55)
  node[above, midway, font=\sffamily\tiny] {$\Delta w$};

\staterow{14}{5.8}{9.4}{7.6}{5.5}{$t_3$}{arb appears!}{1}
\fill[dangerfill, opacity=0.4] (5.5, 13.7) rectangle (5.8, 14.3);
\fill[marketprice] (5.5, 14) circle (3pt);
\draw[arbprofit, thick, <->] (5.5, 13.5) -- (5.8, 13.5);
\node[arbprofit, font=\sffamily\tiny, below=1pt] at (5.65, 13.5) {gap};

\staterow{12}{6.5}{10.1}{8.3}{5.5}{$t_4$}{opportunity grows}{1}
\fill[dangerfill, opacity=0.4] (5.5, 11.7) rectangle (6.5, 12.3);
\fill[marketprice] (5.5, 12) circle (3pt);
\draw[arbprofit, thick, <->] (5.5, 11.5) -- (6.5, 11.5);
\node[arbprofit, font=\sffamily\tiny, below=1pt] at (6.0, 11.5) {larger gap};

\staterow{10}{7.2}{10.8}{9.0}{5.5}{$t_5$}{large opportunity}{1}
\fill[dangerfill, opacity=0.4] (5.5, 9.7) rectangle (7.2, 10.3);
\fill[marketprice] (5.5, 10) circle (3pt);
\draw[arbprofit, very thick, <->] (5.5, 9.5) -- (7.2, 9.5);
\node[arbprofit, font=\sffamily\tiny, below=1pt] at (6.35, 9.5) {large gap};

\draw[arbprofit, line width=2.5pt, ->, rounded corners=5pt]
  (9.0, 9.2) -- (9.0, 8.5) -- (6.3, 8.5) -- (6.3, 8.1);
\node[arbprofit, font=\sffamily\small\bfseries, fill=dangerfill, rounded corners=3pt,
      inner sep=3pt] at (7.65, 8.5) {ARB TRADE};

\staterow{7}{5.5}{9.1}{7.3}{5.5}{$t_5^{+}$}{reset (arb struck)}{0}
\node[marketprice, font=\sffamily\tiny, above=4pt] at (5.5, 7.3) {$m_p$};
\node[poolprice, font=\sffamily\tiny, above=4pt] at (7.3, 7.3) {$m_u$};
\draw[axgray, thick, <->] (5.5, 6.3) -- (7.3, 6.3);
\node[axgray, font=\sffamily\tiny, below=1pt] at (6.4, 6.3) {gap $\approx$ fee};

\staterow{5}{5.8}{9.4}{7.6}{5.5}{$t_6$}{new cycle\ldots}{0}
\draw[weightchange, thick, ->] (6.3, 5.55) -- (6.6, 5.55)
  node[above, midway, font=\sffamily\tiny] {$\Delta w$};

\staterow{3}{6.2}{9.8}{8.0}{5.5}{$t_7$}{building again}{0}
\draw[weightchange, thick, ->] (6.6, 3.55) -- (7.0, 3.55)
  node[above, midway, font=\sffamily\tiny] {$\Delta w$};

\node[axgray, font=\sffamily\Large] at (7.5, 1.8) {$\vdots$};

\draw[axgray, very thick, ->] (0.5, 20.5) -- (0.5, 2.5);
\node[axgray, font=\sffamily\footnotesize, rotate=90, anchor=south] at (0.2, 11.5) {time};

\begin{scope}[yshift=0cm]
  \draw[rounded corners=5pt, thick, axgray, fill=bgfill]
    (1.5, -0.4) rectangle (14.0, 1.2);

  \fill[poolprice] (2.0, 0.6) circle (3pt);
  \node[font=\sffamily\tiny, anchor=west] at (2.3, 0.6) {$m_u$ (implied)};

  \fill[marketprice] (5.0, 0.6) circle (3pt);
  \node[font=\sffamily\tiny, anchor=west] at (5.3, 0.6) {$m_p$ (market)};

  \fill[bandfill] (7.8, 0.4) rectangle (8.4, 0.8);
  \draw[noarb, thick] (7.8, 0.4) rectangle (8.4, 0.8);
  \node[font=\sffamily\tiny, anchor=west] at (8.6, 0.6) {no-arb band};

  \draw[arbprofit, thick, <->] (11.0, 0.55) -- (11.6, 0.55);
  \node[font=\sffamily\tiny, anchor=west] at (11.7, 0.6) {gap};

  \fill[dangerfill, opacity=0.6] (2.0, -0.15) rectangle (2.6, 0.2);
  \node[font=\sffamily\tiny, anchor=west] at (2.8, 0.025) {arb opportunity zone};

  \fill[arbprofit, opacity=0.08] (7.8, -0.15) rectangle (8.4, 0.2);
  \draw[arbprofit, thick, ->] (8.4, 0.025) -- (8.0, 0.025);
  \node[font=\sffamily\tiny, anchor=west] at (8.6, 0.025) {arb trade};
\end{scope}

\end{tikzpicture}
\caption{Complete rebalancing lifecycle showing block-by-block evolution of the no-arb band (green rectangle), implied price $m_u$ (violet dot), and market price $m_p$ (gold dot).
Weights interpolate rightward from $t_0$ to $t_5$, creating a growing gap.
The arb trade at $t_5$ resets $m_u$ so that $m_p$ sits at the band boundary.
At $t_5^+$, a residual gap $\approx$ fee remains.
The cycle restarts from $t_6$.}
\label{fig:lifecycle}
\end{figure}

Figure~\ref{fig:lifecycle} provides a detailed block-by-block view of one complete rebalancing cycle using the number-line representation, combining the band mechanics of Figure~\ref{fig:noarb_schematic} with the auction dynamics of Figure~\ref{fig:auction_schematic}.

Starting from equilibrium at $t_0$ (both $m_u$ and $m_p$ coincide, the band is centred), each block's weight interpolation shifts the band rightward ($t_1$, $t_2$) while the market price $m_p$ stays fixed.
By $t_3$, the market price has fallen outside the left edge of the band and an arb opportunity appears, initially small.
The opportunity grows through $t_4$ and $t_5$ as the band continues its rightward drift.

At $t_5$, an arbitrageur acts.
The trade shifts $m_u$ leftward so that $m_p$ lands at the left boundary of the new, smaller band.
Crucially, the pool does \emph{not} return to perfect alignment ($m_u = m_p$): a residual gap of approximately one fee remains, as explained in Figure~\ref{fig:boundary_targeting}.
From $t_6$ onward, the cycle restarts: the band resumes its rightward drift, the gap begins to build again, and the next arbitrageur waits for the opportunity to exceed their cost.

\end{appendices}

\end{document}